\documentclass{jfm}
\usepackage{changes}
\usepackage{graphicx}
\usepackage{newtxtext}
\usepackage{newtxmath}
\usepackage{natbib}
\usepackage{hyperref}
\usepackage{color}
\usepackage{pdflscape}
\usepackage{afterpage}
\hypersetup{
    colorlinks = true,
    urlcolor   = blue,
    citecolor  = black,
}

\newcommand{\RomanNumeralCaps}[1]

\def\d{\mathrm{d}}

\def\bm{\boldsymbol}
\newcommand{\ed}[1]{\textcolor{red}{#1}}

\title{Addressing bedload flux variability due to grain-shape effects and experimental channel geometry}

\author{Thomas P\"ahtz,\aff{1}
  \corresp{\email{tpaehtz@gmail.com}}
  Yulan Chen,\aff{2}
	Jiafeng Xie,\aff{2,3}
	R\'emi Monthiller,\aff{4}
	Rapha\"el Maurin,\aff{5}
	Katharina Tholen,\aff{6}
	Yen-Cheng Lin,\aff{7}
	Hao-Che Ho,\aff{7}
	Peng Hu,\aff{2}
	\corresp{\email{pengphu@zju.edu.cn}}
	Zhiguo He,\aff{2}
	\corresp{\email{hezhiguo@zju.edu.cn}}
	\and Orencio Dur\'an Vinent\aff{8}}

\affiliation{\aff{1}School of Marine Science and Engineering, South China University of Technology, Guangzhou 511442, China
\aff{2}Institute of Port, Coastal and Offshore Engineering, Ocean College, Zhejiang University, 316021 Zhoushan, China
\aff{3}State Key Laboratory of Fluid Power and Mechatronic Systems and Department of Engineering Mechanics, Zhejiang University, Hangzhou 310027, China
\aff{4}Aix-Marseille University, CNRS, Centrale Marseille, Institut de Recherche sur les Ph\'enom\`enes Hors \'Equilibre (IRPH\'E), Marseille 13384, France
\aff{5}Universit\'e de Toulouse, Toulouse INP, CNRS, IMFT, 31400 Toulouse, France
\aff{6}Institute for Theoretical Physics, Leipzig University, Postfach 100920, 04009 Leipzig, Germany
\aff{7}Department of Civil Engineering, National Taiwan University, Taipei City 106, China
\aff{8}Department of Ocean Engineering, Texas A\&M University, College Station, Texas 77843-3136, USA}

\begin{document}
\maketitle

\begin{abstract}
The study-to-study variability of bedload flux measurements in turbulent sediment transport borders an order of magnitude, even for idealized laboratory conditions. This uncertainty stems from physically poorly supported, empirical methods to account for channel geometry effects in the determination of the transport-driving bed shear stress, and from study-to-study grain-shape variations. Here, we derive a universal method of bed shear stress determination. It consists of a granular-physics-based definition of the bed surface and a channel sidewall correction based on linking Reynolds stress to bulk flow properties via Kolmogorov's theory of turbulence. Application of this method to bedload transport of spherical grains---to rule out grain-shape effects---collapses data from existing laboratory measurements and grain-resolved computation fluid dynamics (CFD) and discrete element method (DEM) simulations for various channel geometries onto a single curve. In contrast, classical sidewall corrections, as well as an alternative bed surface definition, are unable to universally capture these data, especially those from shallow or very narrow channel flows. We then apply our method to an extended grain-shape-controlled data compilation, complemented by literature data for non-spherical grains and from grain-unresolved CFD-DEM simulations. This compilation covers a very diverse range of transport conditions, ranging from very narrow to infinitely wide channels, from shallow to deep channel flows, from mild to steep bed slopes, and from weak to intense transport. We generalize an existing physical bedload flux model to account for grain-shape effects and show that it explains almost all the compiled data within a factor of only $1.3$.
\end{abstract}


\section{Introduction}
Bedload transport is a special kind of sediment transport in which typically coarse sedimentary grains roll, slide, and hop along the surface in response to the shearing of a loose granular bed by a flow of liquid. It plays a vital role in shaping the environments of Earth and other planetary bodies \citep{Poggialietal16} by promoting the formation and growth of geological features of various scales, including ripples and dunes \citep{Charruetal13,Duranetal19}, deltas and fans \citep{Bakeretal15}, and laminations and cross-bedding \citep{Schieber16,Shchepetkinaetal19}. A key problem hampering our understanding of bedload-induced landscape evolution is the notoriously large noise commonly associated with measurements of the bedload flux, often exceeding an order of magnitude \citep{Recking10,AnceyRecking23}. It partially originates from huge non-Gaussian flux fluctuations over large time scales, which even occur under steady flow conditions in the laboratory due to continuous topographic change \citep{DhontAncey18}. However, also when restricted to steady flows over flat beds at short timescales, reported bedload fluxes can still substantially differ between laboratory studies for largely self-similar conditions, for both weak and intense transport-driving flows \citep{AnceyRecking23}, and it has been unclear whether such differences arise due to physical or experimental reasons. On the one hand, physical reasons may come into play because dissimilarities in grain size and shape distributions are not accounted for by commonly used similarity parameters. On the other hand, study-to-study variability of the geometry of laboratory facilities is a potential culprit on the experimental side, especially in view of the associated large random and systematic uncertainties in the experimental determination of the transport-driving bed shear stress $\tau_b$ \citep{Yageretal18b}. Focusing here on rectangular open channels, since they are geometrically similar to natural streams, the ratio $b/h$ between the channel width $b$ and flow depth $h$ is a crucial parameter in this regard. Ideally, one would want $b/h$ to be as large as possible to most closely resemble bedload transport in nature. However, in experimental reality, $b/h$ is often of order unity, substantially weakening the shearing of the bed due to frictional losses at the sidewalls, causing the flow to be nonuniform in the cross-channel direction \citep{Guo15}. Another important parameter is the ratio between $h$ and the grain diameter $d$. For shallow channel flows where $h/d$ is close to unity, the method by which the bed surface elevation, and thus the flow depth $h$ above it, is identified can have a large effect on the value of $\tau_b$.

\begin{figure}
 \centering
 \includegraphics[width=0.5\columnwidth]{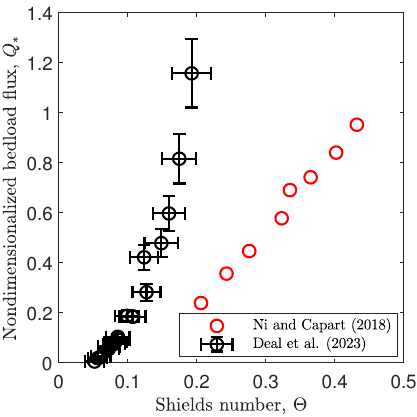}
 \caption{\textbf{Bedload flux measurements for spherical grains by \citet{NiCapart18} and \citet{Dealetal23a}.} The Shields number $\Theta$ and non-dimensionalised bedload flux $Q_\ast$ are defined and determined as described in these studies. For $\Theta\approx0.2$, the values of $Q_\ast$ disagree with each other by a factor of approximately $6$.}
 \label{NiCapartDeal}
\end{figure}

To exemplary illustrate that channel geometry effects often are not appropriately accounted for, we choose two laboratory studies measuring bedload flux over a flat bed: \citet{NiCapart18}, who conducted their experiments in a wide but shallow channel ($h/d\sim2{-}6$), and \citet{Dealetal23a}, who used a very narrow but deep channel ($b/h\approx0.1$). Since the grains in both studies were spherical and relatively uniformly sized, one expects the same quantitative behavior of the non-dimensionalised bedload flux $Q_\ast$ as a function of the non-dimensionalised bed shear stress or Shields number $\Theta$ \citep{Ancey20a}. However, when using these studies' reported values for $\Theta$ and $Q_\ast$, based on their own definitions and methods to determine $\tau_b$, both datasets disagree with each other by a staggering factor of approximately $6$ for $\Theta\approx0.2$ (figure~\ref{NiCapartDeal}).
This indicates that at least one of them was inappropriately correcting for channel geometry effects, showcasing the need for a universal channel geometry correction method.

This study addresses two of the mentioned potential reasons for the study-to-study variability of laboratory bedload flux measurements: grain shape and channel geometry effects. We compile existing and new data of turbulent bedload transport along flat beds (no bedforms) of shape-controlled and (relatively) uniformly sized grains, obtained from experiments and grain-resolved (i.e., directly resolving fluid--particle interactions) and grain-unresolved numerical simulations based on various computational fluid dynamics/discrete element method (CFD-DEM) techniques, across channel flow widths and heights, grain shapes, bed slopes, and flow strengths. In \S\ref{ChannelGeometryCorrectionMethod}, we reconcile channel geometry effects in these data through a universal sidewall correction, based on linking Reynolds stress to bulk flow properties via an extension of Kolmogorov's theory of turbulence \citep{GioiaBombardelli02,GioiaChakraborty06}, and a granular-physics-based definition to precisely determine the bed surface elevation \citep{PahtzDuran18b}. This channel geometry correction method is validated, among others, through a collapse of experimental and grain-resolved numerical bedload flux data for spherical grains on a unique behavior, whereas classical methods \citep{Einstein42,Johnson42,Guo15,Guo17} yield very large scatter. Furthermore, in \S\ref{BedloadModel}, we generalize an existing physical bedload flux model \citep{PahtzDuran20} to account for grain-shape effects. It agrees with the entire channel-geometry-corrected data compilation within a factor of $1.3$. The results are then discussed in \S\ref{Discussion} and conclusions are drawn in \S\ref{Conclusions}.

\section{Universal channel geometry correction method} \label{ChannelGeometryCorrectionMethod}
In this section, we derive and validate a universal channel geometry correction method to determine the transport-driving bed shear stress $\tau_b$ for open channels with hydraulically rough sediment beds and hydraulically smooth sidewalls. We begin with the definition of $\tau_b$ (\S\ref{DefinitionTaub}), and derive from it, partially using known results, an expression that contains the flow depth $h$ and the effective mixture sidewall shear stress $\tau_w$ as unknowns. We then propose to determine $h$ from an existing granular-physics-based definition of the bed surface elevation (\S\ref{BedSurfaceDefinition}), and derive an expression for $\tau_w/\tau_b$ from an extension of Kolmogorov's theory of turbulence (\S\ref{SidewallCorrection}). In \S\ref{Validation}, the resulting channel geometry correction method is tested against a data compilation of turbulent bedload transport of spherical grains and other data.

\subsection{Definition of bed shear stress} \label{DefinitionTaub}
For turbulent bedload transport in open channels with hydraulically rough sediment beds, there is a limited region in the clear-water flow above the bedload layer where the streamwise ($x$) fluid velocity profile $u_x(z)$ approximately obeys the logarithmic law of the wall \citep{Kidanemariam16,Jainetal21,Dealetal23a}:
\begin{equation}
 u_x\simeq\frac{u_\tau}{\kappa}\ln\frac{z+z_o}{z_o}, \label{ulog}
\end{equation}
where $\kappa=0.41$ is the von K\'arm\'an constant, $z$ is the bed-normal coordinate, with $z=0$ the bed surface elevation, $z_o$ is the hydrodynamic roughness, which depends on both grain size and bedload flux \citep{Duranetal12}, and $u_\tau$ is the friction velocity. The latter defines the bed shear stress $\tau_b\equiv\rho_fu_\tau^2$, where $\rho_f$ is the fluid density. In what follows, we seek to express $\tau_b$ in terms of known fluid and channel parameters.

\citet{Kidanemariam16} and \citet{Jainetal21} conducted direct numerical simulations (DNS) with the DEM of bedload transport driven by negative streamwise fluid pressure gradients $\chi$ at zero bed slope angle $\alpha$ for infinitely wide and relatively deep channel flows. Both studies found that $\tau_b$ defined via (\ref{ulog}) is equivalent to the effective mixture shear stress evaluated at the bed surface, $\sigma_{zx}(0)$. Below, we generalize this result to open channels with sidewalls and arbitrary $\alpha$.

For a statistically steady ($\partial_t=0$) and streamwise uniform ($\partial_x=0$) open channel flow, the mixture momentum balance resulting from averaging over time $t$ and the streamwise direction $x$ reads \citep{Pahtzetal26}
\begin{equation}
 \partial_y\tilde\rho\tilde u_y\tilde u_x+\partial_z\tilde\rho\tilde u_z\tilde u_x=\partial_y\tilde\sigma_{yx}+\partial_z\tilde\sigma_{zx}+\chi+\tilde\rho g\sin\alpha, \label{MomMixture}
\end{equation}
where $g$ is the gravitational acceleration, $\tilde\rho$ is the mixture density, $\bm{\tilde u}$ is the mixture velocity, and $\bm{\tilde\sigma}$ is the mixture shear stress tensor, which is composed of the averaged viscous fluid shear stress, the solid phase contact stress, a stress resulting from fluid--particle interactions, and the mixture Reynolds stress. Integrating (\ref{MomMixture}) over height and the cross-channel coordinate $y$, written using the notations $\langle\cdot\rangle_z\equiv\frac{1}{h}\int_0^h\cdot\d z$ and $\langle\cdot\rangle_y\equiv\frac{1}{b}\int_{-b/2}^{b/2}\cdot\d y$, then yields
\begin{equation}
 -b\left[\langle\tilde\sigma_{zx}-\tilde\rho\tilde u_z\tilde u_x\rangle_y\right]^{z=h}_{z=0}=h\left[\langle\tilde\sigma_{yx}-\tilde\rho\tilde u_y\tilde u_x\rangle_z\right]^{y=b/2}_{y=-b/2}+bh\chi+bh\langle\langle\tilde\rho\rangle_y\rangle_zg\sin\alpha. \label{MomMixture2}
\end{equation}
When assuming impermeable boundary conditions at the bed ($\langle\tilde\rho\tilde u_z\rangle_y(0)=0$) and free surface ($\langle\tilde\rho\tilde u_z\rangle_y(h)=0$) on average, then statistical steadiness and mass conservation would imply impermeable boundary conditions also at the sidewalls ($\langle\tilde\rho\tilde u_y\rangle_z(\pm b/2)=0$) on average, and vice versa. Although these boundaries are not necessarily perfectly impermeable, since low vertical transport may occur through the bed surface, we assume impermeability for simplicity, in which case the terms $-\langle\tilde\rho\tilde u_z\tilde u_x\rangle_y(0,h)$ and $-\langle\tilde\rho\tilde u_y\tilde u_x\rangle_z(\pm b/2)$ are Reynolds-like fluctuation stresses. We therefore identify $\sigma_{zx}\equiv\langle\tilde\sigma_{zx}-\tilde\rho\tilde u_z\tilde u_x\rangle_y$, when evaluated at $z=0$ or $z=h$, as the effective mixture bed and free-surface shear stresses, respectively, and $\tau_w\equiv\mp(\langle\tilde\sigma_{yx}-\tilde\rho\tilde u_y\tilde u_x\rangle_z)(\pm b/2)$ as the effective mixture sidewall shear stress. Furthermore, we use the free-surface boundary condition $\sigma_{zx}(h)=0$ and the relation $\langle\tilde\rho\rangle_y=(1-\phi)\rho_f+\phi\rho_p$ \citep{Pahtzetal26}, where $\rho_p$ is the particle density, and $\phi(z)$ is the cross-channel-averaged particle volume fraction, and define the sediment transport load $M\equiv\rho_ph\langle\phi\rangle_z$ (the average mass of transported grains---i.e., grains elevated above the bed surface---per unit area of the bed surface) and the vertical submerged gravitational acceleration $\tilde g_z\equiv(1-\rho_f/\rho_p)g\cos\alpha$. Inserting these relations and definitions, and dividing by $b$, (\ref{MomMixture2}) becomes
\begin{equation}
 \sigma_{zx}(0)=h(\chi+\rho_fg\sin\alpha)-(2h/b)\tau_w+M\tilde g_z\tan\alpha. \label{MomMixture3}
\end{equation}
This expression is a generalization of the mixture balance by \citet{Bagnold56} to the case of a channel with sidewalls ($h/b>0$) and of the clear-water balance by \citet{Guo15} to the case of a mixture ($M>0$); \citet{Guo15} also assumed $\chi=0$. Since we later propose a Bagnoldian-type bedload transport model, we assume analogous to \citet{Bagnold56} that the bed shear stress defined via (\ref{ulog}) is given by
\begin{equation}
 \tau_b=\sigma_{zx}(0)-M\tilde g_z\tan\alpha. \label{taubDefinitionAlt}
\end{equation}
This assumption is the generalization of the mentioned finding $\tau_b=\sigma_{zx}(0)$ for $\alpha=0$ \citep{Kidanemariam16,Jainetal21}. Note that the alternative assumption that $\tau_b=\sigma_{zx}(0)$ for all $\alpha$ would lead to substantial disagreement between the bedload transport model presented in \S\ref{ModelDerivation} and the numerical data against which it is tested. Combining (\ref{MomMixture3}) and (\ref{taubDefinitionAlt}) yields
\begin{equation}
 \tau_b=h(\chi+\rho_fg\sin\alpha)-(2h/b)\tau_w. \label{BedShearStress0}
\end{equation}
Here $\tau_b$ depends on two unknowns: the bed surface elevation $z=0$ required to determine $h$ and the effective mixture sidewall shear stress $\tau_w$. They are addressed below.

\subsection{Definition of bed surface elevation} \label{BedSurfaceDefinition}
The water depth $h$ is the average distance between the free-surface ($z=h$) and bed surface ($z=0$) elevations. Many studies on bedload transport have treated $h$ like it is a well-defined entity, without elaborating on what exactly they mean when referring to the bed surface \citep[e.g.,][]{Dealetal23a}. This non-specificity is acceptable in situations where, regardless of the precise definition of the bed surface elevation $z=0$, the resulting $h$ is much larger than the grain size $d$ \citep[e.g.,][]{Dealetal23a}. This is because distinct definitions of $z=0$ usually differ from one another only by a fraction of $d$, and therefore yield only small uncertainty in the value of $h$ for such deep channel flows due to $h\gg d$. However, for shallow channel flows, the situation is fundamentally different, and a definition that accounts for the relevant grain scale characteristics and/or processes is required.

Most studies have defined $z=0$ via assuming that the relevant grain scale characteristics are the random close packing and near-stationarity of the granular bed underneath $z=0$. For example, $z=0$ has been defined as the elevation where $\phi$ is a constant proportion of the bed packing fraction \citep{Duranetal12,Rebaietal22}, via a yield criterion \citep{Maurinetal15}, via critical values of the average particle velocity $v_x$ \citep{Rebaietal22} or associated shear rate $\dot\gamma=\d_zv_x$ \citep{CapartFraccarollo11}, or a nearly diminishing bedload flux below $z=0$ \citep{NiCapart18}. However, since we later propose a Bagnoldian-type bedload transport model, we are more interested in a definition that accounts for the key assumptions behind such models.

\citet{PahtzDuran18b} demonstrated that there are essentially three assumptions integral to Bagnoldian-type bedload transport models: (i) the vast majority of transport occurs above $z>0$, (ii) the ratio $\mu_b\equiv-\sigma^p_{zx}(0)/\sigma^p_{zz}(0)$ between the bed surface particle shear stress $\sigma^p_{zx}(0)$ and surface particle pressure $-\sigma^p_{zz}(0)$ is approximately constant, and (iii), except for intense transport conditions, the bed surface fluid shear stress $\sigma^f_{zx}(0)$ reduces to approximately the threshold shear stress $\tau_t$. They argued that these assumptions are also expected to be satisfied for the flow-driven saltation of grains along a rigid, rough wall instead of a granular bed and therefore looked for a criterion that characterizes grain rebounds at a rough wall. From drawing an analogy to inclined granular flow, they hypothesized that, close to the wall, $-\sigma^p_{zz}\dot\gamma$, which describes the rate at which the granular flow converts mean particle kinetic energy density into solid phase Reynolds stress $-\rho_p\phi v_zv_x$, should peak. Translating this hypothesis to sediment transport scenarios, they therefore defined the bed surface elevation through
\begin{equation}
 \max(\dot\gamma\Sigma\phi)=(\dot\gamma\Sigma\phi)(0), \label{BedSurface}
\end{equation}
with $\Sigma\phi(z)\equiv\int_z^h\phi(z^\prime)\d z^\prime$, based on the approximation $\sigma^p_{zz}\approx-\rho_p\Sigma\phi\tilde g_z$. They found that in their coupled Reynolds-averaged Navier-Stokes (RANS)-DEM simulations of bedload transport (and windblown sand), when $z=0$ is defined through (\ref{BedSurface}), it, indeed, approximately satisfies the three above assumptions across their simulated transport conditions.

\subsection{Sidewall correction} \label{SidewallCorrection}
Equation~(\ref{BedShearStress0}) is the same as that for clear-water flow \citep{Guo15}, except for generalized meanings of $\tau_b$ and $\tau_w$. In particular, the bed shear stress $\tau_b$, even though it is based on the mixture flow above the bed surface, controls, like its clear-water counterpart, the slope of the logarithmic velocity profile in (\ref{ulog}) \citep{Kidanemariam16,Jainetal21,Dealetal23a}. Like previous studies on the sidewall correction for open channel flows with sediment transport \citep{Einstein42,Johnson42,VanoniBrooks57,Guo15}, we therefore model $\tau_b$ and $\tau_w$ as if the the flow above the bed surface were clear-water flow. 

We start by rearranging (\ref{BedShearStress0}) into the form
\begin{equation}
 \tau_b=\frac{h(\chi+\rho_fg\sin\alpha)}{1+\frac{2h}{b}f_w/f_b}, \label{taubcorr}
\end{equation}
where $f_w\equiv\tau_w/(\rho_fU^2)$ and $f_b\equiv\tau_b/(\rho_fU^2)$ are the friction factors associated with the sidewalls and bed surface, respectively, with $U$ the bulk fluid flow velocity. The rougher these surfaces are, the larger is the rate at which the bulk flow energy density $\rho_fU^2/2$ is frictionally dissipated, and the larger is the shear stress applied at them. The difficulty is now to relate the friction factor ratio $f_w/f_b$ to bulk flow properties, such as the open channel flow geometry and $U$. In fact, one wants to avoid linking $f_w/f_b$ to local flow properties, such the near-bed and near-sidewall flow velocity gradients, since they are usually not readily accessible.

A classical approach employed in numerous studies is to assume $f_w=f_b$, in which case $\tau_b=\tau_R\equiv R(\langle\chi\rangle_z+\rho_fg\sin\alpha)$, with $R\equiv hb/(b+2h)$ the hydraulic radius of the open channel \citep{Guo15}. This prediction is consistent with measurements of centerline flow velocity profiles in particle-free open channels consisting of only hydraulically smooth walls (sidewalls and bottom wall) for a large range of width-to-depth ratios ($b/h=1.04{-}9.5$) \citep{Yangetal05}. However, in bedload transport, where the bed surface is typically hydraulically rougher than the hydraulically smooth sidewalls ($f_b>f_w$), $\tau_R$ merely constitutes a lower bound of $\tau_b$.

Another classical approach is the one by Einstein and Johnson \citep{Einstein42,Johnson42,VanoniBrooks57,Guo17} (see Appendix~\ref{SidewallCorrectionEinstein}), in which $f_w$ is determined from the empirical von K\'arm\'an--Prandtl law as the friction factor of a hydraulically fully smooth channel, and then $f_b$ is linked to $f_w$ via an \textit{ad hoc} separation of the open channel into three parallel, non-interacting sidewall-free channels. However, this separation's only physical justification is its mathematically constructed consistency at the extreme limit $f_b=f_w$, where it rightfully predicts $\tau_b=\tau_R$.

Our approach presented below does not rely on Einstein's \textit{ad hoc} channel separation but instead employs, for the Reynolds shear stress $\tau_s$ at a given surface (the sidewalls or the bed surface), a relation based on Kolmogorov's theory of turbulence \citep{GioiaBombardelli02,GioiaChakraborty06}:
\begin{equation}
 \tau_s=-\rho_f\langle u_\parallel^\prime u_\perp^\prime\rangle=\kappa_\tau\rho_fUu_s, \label{Ansatz}
\end{equation}
where $\langle\cdot\rangle$ denotes an averaging procedure and $\kappa_\tau$ is a flow-geometry-dependent empirical coefficient. Equation~(\ref{Ansatz}) states that the surface-parallel velocity fluctuations $u_\parallel^\prime$ are dominated by the largest turbulent eddies in the system, whose characteristic velocity is $U$, whereas the surface-normal velocity fluctuations $u_\perp^\prime$ are primarily controlled by the largest normal velocities $u_s$ that turbulent eddies can generate near the surface. This relation, which was previously used to physically derive Manning's \citep{GioiaBombardelli02} and Nikuradse's \citep{GioiaChakraborty06} empirical formulas, leads to $f_w/f_b=u_w/u_b$, where $u_b$ and $u_w$ are the values of $u_s$ for the hydraulically rough bed surface and hydraulically smooth sidewalls, respectively.

One may question (\ref{Ansatz}), since it assumes that the sizes of the eddies associated with the bed and those associated with the sidewalls are independent of each other. Einstein's sidewall correction exhibits an analogous characteristic, and \cite{Guo15} therefore challenged it by noting that it would produce a uniform shear stress on all boundaries, $\tau_b=\tau_w=\tau_R$, in hydraulically fully smooth or hydraulically fully rough channels, which should not be the case due to secondary flows. However, we reiterate that, for hydraulically fully smooth channels, $\tau_b=\tau_R$ is well supported by measurements of centerline velocity profiles for a large range of width-to-depth ratios ($b/h=1.04{-}9.5$) \citep{Yangetal05}, and we therefore believe that the assumption of independent eddy sizes is justified, at least in the context of determining $\tau_b$.

The sizes of the eddies governing $u_b$ and $u_w$ are controlled by the typical size $r$ of the protrusion of bed surface grains (which roughly corresponds to the size of the largest eddies that fit into the cavities of the bed surface) and the thickness $5(\kappa_\epsilon\kappa_u^3/2)^{-1/4}R\Rey^{-3/4}$ of the viscous sublayer adjacent to the hydraulically smooth sidewalls \citep{GioiaChakraborty06}, respectively, where $\Rey\equiv UR/\nu$ is the bulk flow Reynolds number, with $\nu$ the kinematic fluid viscosity. Furthermore, $\kappa_\epsilon=5/4$ is a constant that follows from Kolmogorov's four-fifths law, and $\kappa_u\equiv u_R/U$ is a parameter that relates the characteristic velocity $u_R$ of the largest eddies to $U$. It is found to be largely insensitive to flow geometry, with $\kappa_u=0.036\pm0.005$ measured in pipe flow \citep{AntoniaPearson00} and $\kappa_u=0.033$ in the atmospheric boundary layer \citep{TennekesLumley72}. With these parameters, $f_w/f_b=u_w/u_b$ can be calculated via the smooth ($f_w$) and rough ($f_b$) limits of equation~(2) of \citet{GioiaChakraborty06} as
\begin{equation}
 \frac{f_w}{f_b}=2^{1/12}\frac{\sqrt{\Gamma_{-2/3}(\beta/5)}\left[\beta(\kappa_\epsilon\kappa_u^3)^{-1/4}\right]^{1/3}\Rey^{-1/4}}{\sqrt{3/2}(r/R)^{1/3}}=c_{wb}\frac{\Rey^{-1/4}}{(r/R)^{1/3}}, \label{fwfb}
\end{equation}
where $\Gamma_{-2/3}$ is the gamma function of order $-2/3$, $\beta=[3\Gamma(4/3)]^{3/4}$ is the characteristic constant of the exponential decay of the velocity spectrum of turbulent flow, with $\Gamma$ the gamma function, and $c_{wb}\simeq2.23$ is the scaling coefficient resulting from evaluating the expression in the middle of (\ref{fwfb}). Note that the prefactor $2^{1/12}$ originates from the fact that $R$ in \citet{GioiaChakraborty06} refers to the pipe radius, which is twice the hydraulic radius $R$ here. Since $c_{wb}$ depends only on $\kappa_\epsilon$, $\kappa_u$, and $\beta$, but not on the parameter $\kappa_\tau$ defined through (\ref{Ansatz}), $c_{wb}$ is an approximately channel-geometry-independent parameter.

Equation~(\ref{fwfb}) states that $f_w/f_b$ is the ratio between the Blasius friction factor $f_w\sim\Rey^{-1/4}$ for hydraulically fully smooth channels and the friction factor $f_b\sim(r/R)^{1/3}$ for hydraulically fully rough channels. Both scalings are of similar importance for the overall behavior of $f_w/f_b$, and are readily accessible for typical sediment transport datasets (e.g., see Table~\ref{Data} for the values of $\Rey$ and $R/r$ for the channels with sidewalls considered in this paper).
\afterpage{
\clearpage
\begin{landscape}
\begin{table}
 \centering
 \begin{tabular}{l|c|c|c|ccc|ccccccc}
  \hline
  Experimental or numerical study&Width-depth&Friction factor&Grain&\multicolumn{3}{c|}{Shape parameters}&\multicolumn{7}{c}{Hydrodynamic conditions}\\
  &ratio, $b/h$&ratio, $f_w/f_b$&shape&$S_f$&$C_D$&$\mu_s$&$\Rey$&$s$&$Ga$&$\Theta$&$\alpha$&$h^\ast$&$R/r$\\
  \hline
  D23EXPs \citep{Dealetal23a}&$0.09{-}0.16$&$0.31{-}0.36$&spheres&$1$&$0.43$&$0.46$&$3760{-}6557$&2.57&$1334{-}1339$&$0.056{-}0.22$&$0.03{-}0.13$&$13{-}24$&$2.0$\\
  D23EXPe \citep{Dealetal23a}&$0.10{-}0.14$&$0.32{-}0.35$&ellipsoids&$0.83$&$0.60$&$0.60$&$4340{-}5553$&2.42&$1312{-}1314$&$0.071{-}0.16$&$0.03{-}0.09$&$15{-}22$&$2.0$\\
  D23EXPc \citep{Dealetal23a}&$0.10{-}0.13$&$0.36{-}0.39$&chips&$0.51$&$0.39$&$0.65$&$4258{-}5634$&2.36&$767{-}768$&$0.092{-}0.21$&$0.04{-}0.08$&$22{-}29$&$2.8$\\
  D23EXPg \citep{Dealetal23a}&$0.10{-}0.15$&$0.36{-}0.40$&gravel&$0.68$&$0.54$&$0.78$&$4135{-}6344$&2.48&$733{-}735$&$0.086{-}0.26$&$0.04{-}0.11$&$20{-}32$&$2.9{-}3.0$\\
  D23EXPp \citep{Dealetal23a}&$0.09{-}0.12$&$0.38{-}0.41$&prisms&$0.88$&$0.79$&$0.86$&$3769{-}4917$&2.40&$725$&$0.071{-}0.14$&$0.03{-}0.06$&$26{-}35$&$2.9{-}3.0$\\
  NC18EXP \citep{NiCapart18}&$5.2{-}7.5$&$0.39{-}0.42$&spheres&$1$&$0.49$&$0.45$&$4427{-}8568$&$1.39$&$998$&$0.12{-}0.21$&$0.02{-}0.03$&$2.3{-}3.3$&$3.6{-}4.8$\\
  R22EXPc \citep{Rebaietal22}&$2.8{-}4.5$&$0.36{-}0.75$&cylinders&$0.84$&$0.32$&$0.60$&$3282{-}39287$&$1.42$&$738$&$0.25{-}1.06$&$0.05{-}0.06$&$8.7{-}14$&$12{-}16$\\
  R22EXPl \citep{Rebaietal22}&$5.0{-}6.0$&$0.57{-}0.74$&lenses&$0.55$&$0.45$&$0.62$&$6110{-}16151$&$1.37$&$196$&$0.99{-}1.65$&$0.03{-}0.05$&$15{-}18$&$23{-}26$\\
  \hline
  Z22LESn \citep{Zhangetal22}&$0.13$&$0.29^\ast$&spheres&$1$&0.48&0.45&N/A&$2.55$&$1354{-}1356$&$0.11{-}0.20$&$0.06{-}0.11$&$16{-}17$&$2.0$\\
  Z25LESn \citep{Zhangetal25}&$0.12$&$0.29^\ast$&gravel&$0.67$&0.64&0.75&N/A&$2.47$&$615{-}617$&$0.12{-}0.36$&$0.03{-}0.10$&$29$&$3.3$\\
  Z22LESw \citep{Zhangetal22}&$\infty$&N/A&spheres&$1$&0.48&0.45&N/A&$2.55$&$1378$&$0.047{-}0.14$&$0.01{-}0.03$&$4.6{-}22$&$9.1{-}43$\\
	Z25LESw \citep{Zhangetal25}&$\infty$&N/A&gravel&$0.67$&0.64&0.75&N/A&$2.47$&$617$&$0.18{-}0.29$&$0.01{-}0.014$&$29{-}30$&$59{-}60$\\
  J21DNS \citep{Jainetal21}&$\infty$&N/A&spheres&$1$&N/A&N/A&N/A&$2.55$&$44.7$&$0.13$&$0$ ($\chi>0$)&$18$&$36$\\
  KU17DNS&$\infty$&N/A&spheres&$1$&N/A&N/A&N/A&$2.5$&$28.37$&$0.096{-}0.19$&$0$ ($\chi>0$)&$13$&$26$\\
  \hline
  M18RANS \citep{Maurinetal18}&$\infty$&N/A&spheres&$1$&$0.42$&$0.40$&N/A&$1.75{-}4$&$1259{-}2521$&$0.11{-}1.38$&$0.01{-}0.19$&$2.0{-}62$&$4.0{-}123$\\
  M19RANSs \citep{Monthiller19}&$\infty$&N/A&spheres&$1$&$0.42$&$0.49$&N/A&$2.5$&$1782$&$0.16{-}0.72$&$0.05$&$4.1{-}18$&$8.3{-}37$\\
  M19RANSt1 \citep{Monthiller19}&$\infty$&N/A&triplets&$0.85$&$0.42$&$0.55$&N/A&$2.5$&$1782$&$0.24{-}1.01$&$0.05$&$6.1{-}26$&$12{-}51$\\
  M19RANSt2 \citep{Monthiller19}&$\infty$&N/A&triplets&$0.75$&$0.42$&$0.63$&N/A&$2.5$&$1782$&$0.31{-}1.28$&$0.05$&$8.0{-}33$&$16{-}65$\\
  M19RANSt3 \citep{Monthiller19}&$\infty$&N/A&triplets&$0.71$&$0.42$&$0.68$&N/A&$2.5$&$1782$&$0.39{-}1.57$&$0.05$&$9.9{-}40$&$20{-}80$\\
  M19RANSt4 \citep{Monthiller19}&$\infty$&N/A&triplets&$0.62$&$0.42$&$0.72$&N/A&$2.5$&$1782$&$0.47{-}1.86$&$0.05$&$12{-}47$&$24{-}95$\\
  M19RANSt5 \citep{Monthiller19}&$\infty$&N/A&triplets&$0.58$&$0.42$&$0.69$&N/A&$2.5$&$1782$&$0.56{-}2.18$&$0.05$&$14{-}55$&$28{-}111$\\
  LES&$\infty$&N/A&spheres&$1$&$0.54$&$0.49$&N/A&$2.56$&$1341$&$0.14{-}0.32$&$0$ ($\chi>0$)&$18{-}22$&$35{-}44$\\
  LESCD&$\infty$&N/A&spheres&$1$&$1.78$&$0.49$&N/A&$2.56$&$1341$&$0.17{-}0.26$&$0$ ($\chi>0$)&$20{-}23$&$40{-}47$\\
  BLRANS1&$\infty$&N/A&spheres&$1$&$2.24$&$0.38$&N/A&$[1.1,2.65]$&$50$&$0.044{-}1.21$&$0{-}0.16$&$\infty$&$\infty$\\
  BLRANS2&$\infty$&N/A&spheres&$1$&$1.46$&$0.38$&N/A&$[1.1,2.65]$&$100$&$0.058{-}1.21$&$0$&$\infty$&$\infty$\\
  BLRANSCD&$\infty$&N/A&spheres&$1$&$15.9$&$0.38$&N/A&$2.65$&$50$&$0.063{-}0.45$&$0$&$\infty$&$\infty$\\
  \hline
 \end{tabular}
 \caption{{\bf Summary of bedload flux data shown in this study.} The top, middle, and bottom entries correspond to experiments, grain-resolved CFD-DEM simulations, and grain-unresolved CFD-DEM simulations, respectively. Those data of the datasets BLRANS1 and BLRANS2 with $s=2.65$ and $\alpha=0$ are from \citet{PahtzDuran20}. The datasets Z25LESn and Z25LESw do not include data from simulations based on the `artificial-shrinkage method' by \citet{Zhangetal25}, since this method has been falsified \citep{Chenetal26}. The asterisk $^\ast$ indicates that the friction factor ratio $f_w/f_b$ was not calculated but determined empirically. Note that the here listed parameter values of $C_D$, $Ga$, $\Theta$, and $h^\ast$ are based on $d=c_g$.}
 \label{Data}
\end{table}
\end{landscape}
}
They require knowledge of the bulk fluid flow velocity $U$ and bed grain protrusion size $r$, respectively. We calculate $U$ from the fluid discharge $Q_f$ as \citep{Rebaietal22}
\begin{equation}
	U=\frac{Q_f}{b\left(h-\int_0^h\phi\d z\right)}. \label{U}
\end{equation}
Note that, except for intense bedload transport, the term $\int_0^h\phi\d z$ has only a marginal effect on $U$. Hence, for the datasets by \citet{Dealetal23a}, which will constitute a part of our data compilation and for which concentration profile data are not available, we calculate $U$ as $U\simeq Q_f/(bh)$.

Furthermore, for random loose packed beds, like those in the classical experiments by \citet{Nikuradse33}, where many bed surface grains protrude nearly fully into the flow, $r$ is approximately equal to the size $k$ of the physical roughness elements \citep{GioiaBombardelli02,GioiaChakraborty06} and may be approximated as the volume-equivalent sphere diameter $d_o\equiv\sqrt[3]{6V_p/\pi}$, where $V_p$ is the grain volume. However, in the context of bedload transport over mobile beds, the action of the flow smoothes out and compactifies the bed surface \citep{AllenKudrolli18,Cunezetal22}, since flow-mobilized bed surface grains tend to resettle in more resistant bed pockets \citep{Clarketal17}. As a result, the longer a bed is exposed to flow, the less do grains tend to protrude into it, and this process continues until a state of maximum resistance is reached \citep{Pahtzetal20a}. We estimate $r$ in this state from laminar bedload experiments with spherical grains \citep{Charruetal04,Ouriemietal07,Houssaisetal15,Cunezetal22}. They suggest that a Shields number $\Theta\approx0.1$ is needed to mobilize grains in the state of maximal resistance, which we use to infer the typical pivoting angle of grains resting in bed pockets as $\psi\approx60^\circ$ from figure~13 of \citet{Agudoetal17}. It corresponds to a protrusion of $r=d\cos\psi\approx d/2$.

Moreover, for non-spherical grains, we exploit that grains transported at elevations close to the bed surface tend to align their largest projected area parallel to the bed due to torque \citep{Jainetal20,Zhangetal25}. We therefore choose the shortest grain axis $c_g$ as the parameter setting $r$. It complements the large ($a_g$) and intermediate ($b_g$) grain axes, defined through approximating grains as ellipsoids. Hence, we obtain $r=c_g/2$ as an estimate for sediment transport applications. However, when comparing to data obtained from open-channel flow over manually prepared fixed beds \citep{Song94,SongChiew01} or artificial surfaces \citep{Aueletal14}, we use the standard $r=d_o$ or $r=k$, respectively.

\subsection{Validation of the channel geometry correction method} \label{Validation}
\subsubsection{Data compilation for turbulent bedload transport of spherical grains} \label{DataCompilationSpherical}
In order to isolate channel geometry effects on bedload flux, we compile existing experimental datasets for \textit{spherical grains}, assuming that grain-shape effects are negligible for them. However, surprisingly, even though turbulent bedload transport has been studied for more than a century \citep{Ancey20a}, flux measurements for spherical grains with a nearly uniform size distribution seem to be exceedingly scarce. (By contrast, such kinds of measurements are the rule rather than exception when studying bedload transport driven by laminar flow; e.g. \citet{Ouriemietal07,Aussillousetal13,Houssaisetal15,Cunezetal22}.) After an extensive literature search, we have managed to find only a handful of datasets \citep{NiCapart18,Freyetal06,Frey14,ArmaniniCavedon19,Carrilloetal21,Dealetal23a}. Four of these were subsequently disregarded because of either partially crystallised beds and very large scatter \citep{Freyetal06,Frey14}, or the presence of substantial suspended load in addition to bedload \citep{ArmaniniCavedon19}, or self-inconsistent data \citep{Carrilloetal21} (almost no bedload flux change with important variation of flow strength: increasing bed slope at constant flow depth). The only two remaining experimental datasets for spherical grains are those by \citet{Dealetal23a} and \citet{NiCapart18} discussed in the Introduction. We therefore supplement the data compilation with additional datasets from numerical simulations that couple the DEM for the particle phase with a CFD method for the fluid phase that resolves the sub-grain scale: two datasets based on Direct Numerical Simulations (DNS) \citep{Jainetal21,KidanemariamUhlmann17} for infinitely wide channels ($b/h=\infty$, i.e., periodic boundaries), and two datasets based on grain-resolved large-eddy simulations (LES) \citep{Zhangetal22}: one for $b/h=\infty$ and one for a narrow-deep-channel configuration ($b/h\approx0.1$). Only such high-end methods, because they do not coarse-grain the particle phase and do not rely on empiricism for fluid--particle interactions, can be considered as similarly reliable as controlled experiments. The above spherical-grain datasets are denoted as D23EXPs, NC18EXP, Z22LESn, Z22LESw, J21DNS, and KU17DNS, and summarized in Table~\ref{Data} together with other datasets used later. Note that, to Z22LESn and Z25LESn (used later), we are unable to apply our sidewall correction method, since the bulk flow velocity $U$, or data from which it can be inferred, was not reported. In addition, wall-modeled LES simulations such as these are known to underestimate the friction factors of channels with smooth walls \citep{Nikitinetal00,Yangetal17}, rendering (\ref{fwfb}) unusable. We therefore use the empirical sidewall correction that \citet{Zhangetal25} applied to Z22LESn and Z25LESn, $\tau_b=3\tau_R$, corresponding to $f_w/f_b\simeq0.29$, which is justified because it leads to an overlap with the sidewall-free simulations Z22LESw and Z25LESw, respectively, for otherwise almost the same conditions (Table~\ref{Data}).

\begin{figure}
 \centering
 \includegraphics[width=0.5\columnwidth]{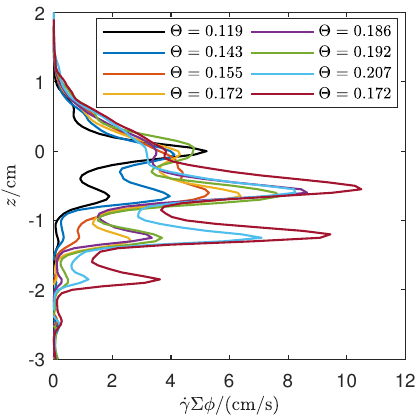}
 \caption{\textbf{Determination of bed surface elevation for experimental dataset NC18EXP \citep{NiCapart18}.} The vertical profiles of $\dot\gamma\Sigma\phi$ exhibit several local maxima due to layering. We choose the top-most local maximum as the bed surface elevation, since the vertical profile of the fluid shear stress $\sigma^f_{zx}(z)$ exhibits a focal point there, $\sigma^f_{zx}(0)\approx\tau_t$ (\S\ref{FurtherTests}), consistent with the theoretical expectation \citep{PahtzDuran18b}.}
 \label{BedSurfaceDetermination0}
\end{figure}

For the spherical-grain datasets discussed in the previous paragraph, we determine the bed surface elevation using the method described in \S\ref{BedSurfaceDefinition}, provided that the required data are available from the studies. This is the case for the dataset NC18EXP (figure~\ref{BedSurfaceDetermination0}), but not for D23EXPs, Z22LESn, Z22LESw, J21DNS, and KU17DNS.
For the latter, we use the values reported in the respective studies. This approximation is justified for D23EXPs, Z22LESn, J21DNS, and KU17DNS, since $h^\ast=h/d\gg1$ (Table~\ref{Data}). In the case of Z22LESw, the authors of the respective study \citep{Zhangetal22} stated that they employed the method by \citet{PahtzDuran18a} to calculate the bed surface elevation, which is the same as our method described in \S\ref{BedSurfaceDefinition}. For more details and the calculation of the bed shear stress uncertainty, see Appendix~\ref{BedShearStressEstimation}.

\subsubsection{Test of channel geometry correction methods against data compilation}
\begin{figure}
 \centering
 \includegraphics[width=1.0\columnwidth]{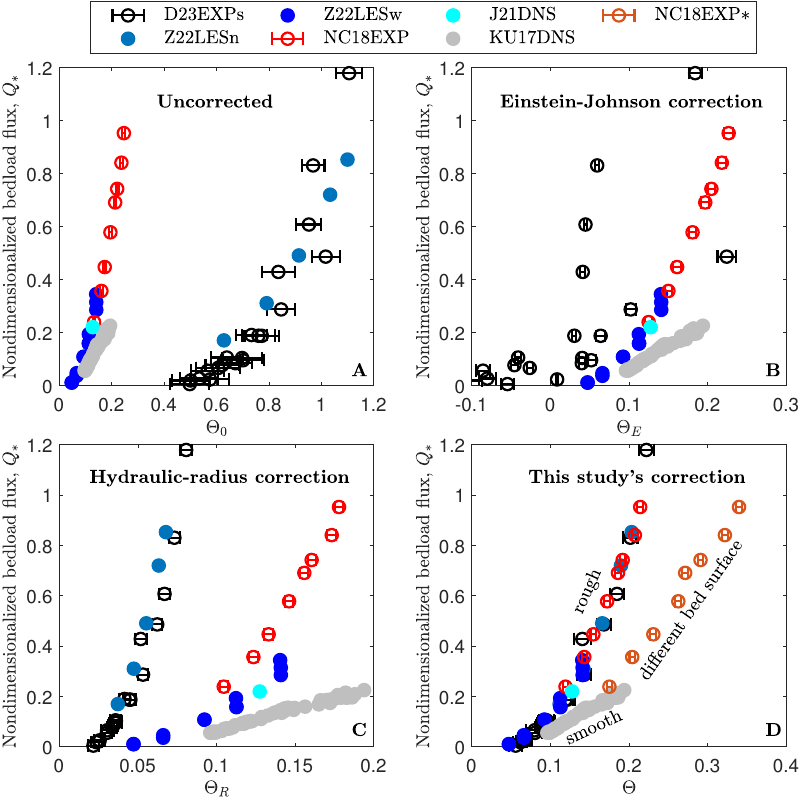}
 \caption{\textbf{Channel geometry corrections.} Non-dimensionalised bedload flux $Q_\ast$ versus Shields numbers for spherical grains: (A) uncorrected for sidewall effects ($\Theta_0$); (B) hydraulic-radius correction \citep{Guo15} ($\Theta_R$); (C) Einstein--Johnson correction \citep{Einstein42,Johnson42,VanoniBrooks57,Guo17} ($\Theta_E$); (D) our universal correction ($\Theta$). Symbols correspond to data from experiments and grain-resolved simulations (Table~\ref{Data}). An error bar indicates the standard error and/or uncertainty range. In the absence of error bars, uncertainties are smaller than the symbol size. Except for KU17DNS ($s^{1/2}Ga\simeq44.86$, smooth), all data are in the rough transport regime ($s^{1/2}Ga>70$). The values of $h$ used to determine $\tau_b$ in Z22LESw and NC18EXP use a granular-physics-based bed surface definition \citep{PahtzDuran18b}, (\ref{BedSurface}), and in NC18EXP$\ast$ are based on the original definition reported by \citet{NiCapart18}. For all other datasets, reported values of $h$ are used as it has a negligible effect on the Shields numbers due to $h\gg d$. Dataset Z22LESn in D is based not on our sidewall correction, but on the correction that \citet{Zhangetal25} applied to this data set, $\Theta=3\Theta_R$ (see text).}
 \label{SphereData}
\end{figure}

In this subsubsection, we test our and existing channel geometry correction methods against the spherical-grain data compilation discussed above. Once the bed shear stress $\tau_b$ is determined using a given channel geometry correction method, we are able to fully characterize the hydrodynamic bedload transport conditions by dimensionless similarity parameters. We choose the particle--fluid density ratio $s\equiv\rho_p/\rho_f$, Galileo number $Ga\equiv d\sqrt{s\tilde g_zd}/\nu$, Shields number $\Theta\equiv\tau_b/(\rho_p\tilde g_zd)$, dimensionless flow depth $h^\ast\equiv h/d$, and bed slope angle $\alpha$ (Table~\ref{Data}). All spherical-grain datasets (and also those for non-spherical grains considered later) are well within the range of what is typically categorized as bedload transport \citep[$s\lesssim10$;][]{PahtzDuran20}, and most of them also satisfy $s^{1/2}Ga\gtrsim70$, here termed ``rough'' transport conditions. In rough turbulent bedload, particle inertia dominates viscous fluid--particle interactions, and the non-dimensionalised bedload flux $Q_\ast\equiv Q/(\rho_pd\sqrt{s\tilde g_zd})$ is therefore expected to become essentially independent of $s$ and $Ga$ for sufficiently small $\alpha$ \citep{Ancey20a,PahtzDuran20,Zhangetal22}. We use this expectation as a tool to evaluate the different sidewall corrections outlined above (figure~\ref{SphereData}).
It can be seen that the data for rough turbulent bedload transport of spherical grains, indeed, collapse, within measurement uncertainty, on a universal $Q_\ast(\Theta)$-behavior for our method (figure~\ref{SphereData}D), but do not collapse at all when employing a classical method or left uncorrected (figures~\ref{SphereData}A--C), or when using our sidewall correction but a different bed surface definition (NC18EXP$\ast$ in figure~\ref{SphereData}D). Note that the spherical-grain dataset Z22LESn is not included in figure~\ref{SphereData}C, since the data for the bulk flow velocity $U$, required to calculate the Einstein--Johnson sidewall correction, are unavailable. Furthermore, Z22LESn in figure~\ref{SphereData}D is based not on our sidewall correction (since it also requires $U$ as an input), but on the correction that \citet{Zhangetal25} applied to this data set, $\tau_b=3\tau_R$.

\subsubsection{Further independent tests of our channel geometry correction method} \label{FurtherTests}
\begin{figure}
 \centering
 \includegraphics[width=1.0\columnwidth]{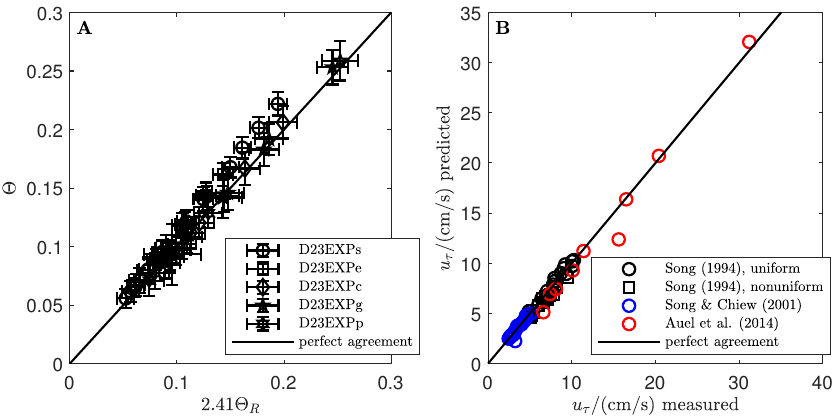}
 \caption{\textbf{Channel geometry corrections.} (A) Plot of $\Theta$ versus $2.41\Theta_R$ (using $d=c_g$ for non-spherical-grain data, as explained in \S\ref{BedloadFluxModel}), based on $\tau_b=2.41\tau_R$, the empirical correction in \citet{Dealetal23a}. (B) Log-profile shear velocity $u_\tau\simeq\sqrt{\tau_b/\rho_f}$ predicted by our sidewall correction versus measured one for experimental data from particle-free flows in open channels consisting of artificially created hydraulically rough beds ($r=d_0$ or $r=k$) and hydraulically smooth sidewalls \citep{Song94,SongChiew01,Aueletal14} (data as summarised in the tables of \citet{Guo15}). For nonuniform flows, the bed slope $\tan\alpha$ is corrected using equation~(1b) of \citet{Guo15}.}
 \label{SphereData2}
\end{figure}

Figures~\ref{SphereData2}A and \ref{SphereData2}B present further independent support for our sidewall correction. First, our correction predicts almost the same Shields numbers as those that \citet{Dealetal23a} empirically determined as $\tau_b=2.41\tau_R$ via fitting (\ref{ulog}) to their flow velocity data, for both spherical and non-spherical grains (figure~\ref{SphereData2}A).
Second, it is also consistent with experimental data from particle-free flows in open channels ($b/h=2.94{-}7.79$) consisting of hydraulically rough beds and hydraulically smooth sidewalls \citep{Song94,SongChiew01,Aueletal14} (figure~\ref{SphereData2}B).

\begin{figure}
 \centering
 \includegraphics[width=0.5\columnwidth]{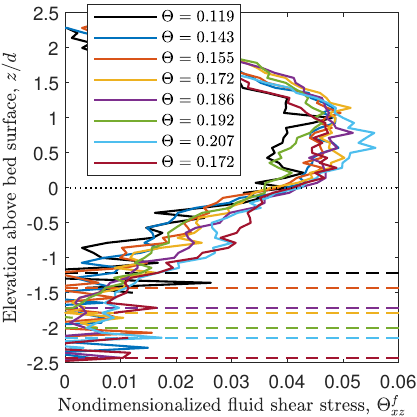}
 \caption{\textbf{Independent evidence supporting our bed surface elevation definition.} Vertical profiles (solid lines) of the non-dimensionalised fluid shear stress $\Theta^f_{zx}\equiv\sigma^f_{zx}/(\rho_p\tilde g_zd)$ for the dataset NC18EXP \citep{NiCapart18}. The data exhibit a focal point at the bed surface elevation $z=0$ defined by (\ref{BedSurface}): $\Theta^f_{zx}(0)\approx0.04$, a property required for Bagnoldian-type bedload models \citep{PahtzDuran18b}. The fluid shear stress $\sigma^f_{zx}$ is calculated from local quantities at the channel centre as described in \citet{Chauchat18}, assuming that the mixture viscosity obeys the closure measured by \citet{Boyeretal11} for viscous suspensions. The dashed lines correspond to the bed surface elevations determined by \citet{NiCapart18}.}
 \label{BedSurfaceDeterminationSupport}
\end{figure}

Furthermore, the strong quantitative difference between the spherical-grain data by \citet{Dealetal23a} and \citet{NiCapart18} discussed in the Introduction (figure~\ref{NiCapartDeal}) is found to be predominantly the consequence of the latter authors' bed surface definition (NC18EXP$\ast$ in figure~\ref{SphereData}D). In fact, figure~\ref{BedSurfaceDeterminationSupport} shows that the bed surface elevations resulting from their definition are much below ours.
It can also be seen that the vertical profiles of the non-dimensionalised local fluid shear stress $\Theta^f_{zx}$ for various flow strengths exhibit a focal point at the bed surface elevation ($z=0$) determined through our method, (\ref{BedSurface}), with a focal value $\Theta^f_{zx}(0)\approx0.04$ close to the transport threshold Shields number $\Theta_t$. As discussed in \S\ref{BedSurfaceDefinition}, the occurrence of such a focal point is precisely one of the properties that we expect the bed surface to have.

\section{Physical rough-bedload flux model across grain shapes, bed slopes, flow strengths and depths} \label{BedloadModel}
In this section, we derive (\S\ref{ModelDerivation}) and validate (\S\ref{ModelValidation}) a general bedload flux model. The validation makes use of an extended data compilation that complements the compilation for spherical grains, used in \S\ref{ChannelGeometryCorrectionMethod} to test channel geometry correction methods, with existing data from grain-shape-controlled experiments and grain-resolved CFD-DEM simulations for non-spherical grains, and with data for spherical and non-spherical grains from grain-unresolved CFD-DEM simulations (\S\ref{DataCompilationExtended}).

\subsection{Model derivation} \label{ModelDerivation}
The model is derived in two steps. First, in \S\ref{BedloadFluxModel}, via generalizing an existing bedload flux model \citep{PahtzDuran20}, we obtain an expression that predicts the non-dimensionalised bedload flux $Q_\ast$ as a function of the excess Shields number $\Theta-\Theta_t$. Then, in \S\ref{TransportThreshold}, we explain how we calculate the unknown transport threshold $\Theta_t$ appearing in this expression.

\subsubsection{Bedload flux} \label{BedloadFluxModel}
Using a numerical model coupling the RANS equations with the DEM, \citet{PahtzDuran20} simulated non-suspended sediment transport of spherical grains and derived a general transport rate expression holding for a large range of conditions, including turbulent bedload with $s^{1/2}Ga\gtrsim80$ and windblown sand:
\begin{equation}
	Q_\ast=2\kappa^{-1}\sqrt{\Theta_t}M_\ast(1+c_MM_\ast), \label{Q}
\end{equation}
with $M_\ast\equiv M/(\rho_pd)$. Based on the definition $M\equiv\rho_ph\langle\phi\rangle_z$ of the transport load in \S\ref{DefinitionTaub}, $M_\ast$ can be interpreted as the number $N$ of transported grains above a bed surface area of size $\sim d^2$. That is, $M_\ast$ is a measure for the particle activity and therefore proportional to the excess Shields number $\Theta-\Theta_t$ \citep{Ancey20a}; the proportionality factor is derived shortly. Hence, for bedload transport, where $1+c_MM_\ast\approx2\sqrt{c_MM_\ast}$ (arithmetic mean $\approx$ geometric mean) is a good approximation as typically $c_MM_\ast\sim O(1)$, (\ref{Q}) predicts a bedload flux scaling similar to the classical formula by \citet{MeyerPeterMuller48}.

In (\ref{Q}), $2\kappa^{-1}\sqrt{\Theta_t}$ represents the dimensionless average particle velocity at $\Theta_t$, deduced from assuming that $\Theta_t$ corresponds to the smallest Shields number at which a quasi-continuous hopping-rebound motion (saltation) of a test grain along a stationary bed in a grain-motion-undisturbed logarithmic fluid velocity profile can be sustained, a so-called ``rebound threshold'' \citep{Pahtzetal20a}. Furthermore, the term $c_MM_\ast$, with $c_M=1.7$ a constant parameter determined from the simulations, encodes the rate of collisions between transported grains ($\sim M_\ast^2$) relative to that of grain-bed collisions ($\sim M_\ast^1$). In fact, (\ref{Q}) expresses the balance between global energy production due to drag (left-hand side) and energy dissipation due to grain contacts (right-hand side). In order to apply it to non-spherical grains, it is important to generalize $d$ in a manner that retains $M_\ast\sim N$. Since grains transported at elevations close to the bed surface, where most transport takes place, tend to align their largest projected area $A^\mathrm{max}_p$ parallel to the bed due to torque \citep{Jainetal20,Zhangetal25}, implying $M\sim N\rho_pV_p/A^\mathrm{max}_p$, this requirement begets $d\equiv\frac{3}{2}V_p/A^\mathrm{max}_p$. From approximating grains as ellipsoids, we estimate $V_p=\pi a_gb_gc_g/6$ and $A^\mathrm{max}_p=\pi a_gb_g/4$, leading to $d=c_g$ (already used in figure~\ref{SphereData2}A).

In steady, homogeneous sediment transport, the average vertical force acting on transported grains is dominated by gravity, buoyancy, and potentially fluid lift, since drag forces during the grains' upward motion tend to cancel those during their downward motion \citep{PahtzDuran18a}. When fluid lift is neglected, as in the simulations by \citet{PahtzDuran18b,PahtzDuran20}, this implies that the particle pressure at the bed surface $-\sigma^p_{zz}(0)$ is approximately equal to the submerged particle weight $M\tilde g_z$. The bed friction coefficient $\mu_b\equiv-\sigma^p_{zx}(0)/\sigma^p_{zz}(0)$ then links $M\tilde g_z$ to the surface particle shear stress $\sigma^p_{zx}(0)$. When further separating $\sigma^p_{zx}(0)$ into gravity and drag contributions, and assuming that particle drag on the fluid reduces the surface fluid shear stress $\sigma^f_{zx}(0)$ to a value that is just sufficient to sustain transport, one obtains (after some rearrangements) Bagnold's famous expression \citep{Bagnold56,PahtzDuran18b}
\begin{equation}
 M_\ast=\frac{\Theta-\Theta_t}{\mu_b-\tan\alpha}, \label{M0}
\end{equation}
which is a key ingredient of numerous bedload transport and windblown sand models in the literature, including the one by \citet{PahtzDuran20}. Historically, it has been assumed that $\mu_b$ is equal to $\mu_s$, the tangent of the static angle of repose of the bulk bed material \citep{Bagnold56}. However, RANS-DEM simulations of turbulent bedload transport of spherical grains suggest that, although constant, $\mu_b$ is actually substantially larger than $\mu_s$ ($\mu_b\approx0.7$ versus $\mu_s\approx0.4$) \citep{PahtzDuran18b}. This observation can be understood as a bed surface strengthening effect \citep{Clarketal17,AllenKudrolli18,Cunezetal22}. Bedload transport increases the bed surface's ability to resist shear stress, since temporarily mobilized grains can settle again in more stable bed surface pockets \citep{Clarketal17}. Whenever this happens, the bed surface becomes stronger relative to the bulk, i.e., $r_b\equiv\mu_b/\mu_s$ increases. This process continues until a state is reached at which the pockets that emerge at the places where grains are mobilized are equivalent to the most stable pockets that transported grains can reach \citep{Clarketal17}. Then, $r_b$ has acquired its equilibrium value and will no longer increase. In viscous bedload transport, where all pockets are reachable as a transported grain's kinetic energy is typically much smaller than the potential energy wells of the bed surface, $r_b\approx3.4$ \citep{Pahtzetal21}. However, in this study on turbulent bedload transport, where these two energy scales are of comparable size, $r_b$ attains a smaller value as grains are too fast to probe the entire phase space of bed surface pockets \citep{Clarketal17}. We use the value $r_b=1.8$ suggested by the simulations by \citet{PahtzDuran18b}, where $\mu_b\approx0.7$ and $\mu_s\approx0.4$, to calculate $\mu_b=r_b\mu_s$ from the measured $\mu_s$.

In addition to using $\mu_b=r_b\mu_s$, we also generalize (\ref{M0}) to account for lift forces, defined as non-buoyancy fluid forces that act in the bed-normal direction. Neglecting particle-sidewall interactions in the particle momentum balance, a derivation analogous to that in \citet{PahtzDuran20} yields (Appendix~\ref{TransportLoadExpression})
\begin{equation}
 M_\ast=(\Theta-\Theta_t)/\mu^\dagger,\quad\text{with}\quad\mu^\dagger\equiv r_b\mu_s\left[\frac{1+\left(\frac{s\tan\alpha}{s-1}+\chi^\ast\right)\overline{f_L}/\overline{f_D}}{1+r_b\mu_s\overline{f_L}/\overline{f_D}}\right]-\tan\alpha, \label{M}
\end{equation}
where $\chi^\ast\equiv\chi/(\rho_p\tilde g_z)$, $f_D$ is the drag force per unit mass, $f_L$ is the lift force per unit mass divided by the fluid volume fraction $1-\phi$, and the overbar, $\overline{\cdot}\equiv\langle\phi\cdot\rangle_z/\langle\phi\rangle_z$, denotes the $\phi$-weighted height average, i.e. the average over all transported grains. The expression for $\mu^\dagger$ takes into account that Reynolds stress gradients do not contribute to the buoyancy force \citep{Maurinetal18,Zhuetal26}, a recent finding that implies that there is no buoyancy force counteracting the gravitational and pressure gradient forces in the streamwise direction.

Equation~(\ref{M}) contains the single additional calibratable parameter $\overline{f_L}/\overline{f_D}$, and reduces to (\ref{M0}) if fluid lift is neglected. In real-world scenarios, one may expect $\overline{f_L}/\overline{f_D}$ to be of order unity, as suggested by measurements of drag and lift forces acting on surface roughness elements, including grains in bed surface pockets \citep{Chepil58,Chepil61}, and by DNS-DEM simulations of sheared mobile beds \citep{Jietal13}. We use $\overline{f_L}/\overline{f_D}=1.0$ because it optimizes the agreement between the final model and the entire data compilation.

\subsubsection{Transport threshold} \label{TransportThreshold}
We reiterate that the prefactor $2\sqrt{\Theta_t}/\kappa$ in (\ref{Q}) resulted from assuming that $\Theta_t$ is the smallest Shields number that permits a quasi-continuous hopping-rebound motion (saltation) of a test grain along a stationary bed in a grain-motion-undisturbed logarithmic fluid velocity profile \citep{PahtzDuran20}. In such a scenario, lift forces acting on the moving test grain are typically much smaller than drag forces due to the formers' rapid decrease with distance from the bed \citep{Chepil61,Lietal19}. Then, it can be shown that, for conditions with $s^{1/4}Ga\gtrsim200$, $\Theta_t$ in (\ref{Q}) obeys \citep{Pahtzetal21}
\begin{equation}
	\Theta_t=c_t\left(r_b\mu_s-\frac{s}{s-1}\tan\alpha-\chi^\ast\right)/C_D, \label{Thetat}
\end{equation}
where $c_t$ is a dimensionless parameter, and $C_D$ is an effective drag coefficient parametrizing streamwise drag force acting on grains. However, in spite of the similarity of (\ref{Thetat}) with scaling laws classically associated with the incipient motion of a grain resting in a bed surface pocket, we emphasize again that $\Theta_t$ is not an incipient motion threshold but a so-called rebound threshold, as defined and discussed by \citet{Pahtzetal20a} and \citet{Pahtzetal21}. In particular, incipient motion thresholds strongly depend on turbulent fluctuations and therefore exhibit an additional dependence on $h^\ast$ that is responsible for their often observed increase with bed slope $\tan\alpha$ \citep{Lambetal08}. By contrast, the rebound threshold $\Theta_t$ is insensitive to turbulent fluctuations \citep{Pahtzetal20a}, as was also directly shown in figure~S1 of \citet{PahtzDuran18a} based on numerical simulation data by \citet{Maurinetal15}, and therefore always decreases with $\tan\alpha$.

The drag coefficient $C_D$ in (\ref{Thetat}) is smaller than that for settling grains, $C_{D\mathrm{Settle}}\equiv4(s-1)gd/(3v_s^2)$ \citep{Dealetal23a}, where $v_s$ is the terminal grain settling velocity, due to the smaller projected grain area exposed to the flow. Assuming again that transported grains tend to align their largest projected area $A^\mathrm{max}_p=\pi a_gb_g/4$ parallel to the bed, the effective projected area $A^\mathrm{eff}_p$ in the streamwise direction is estimated as the geometric mean of the remaining two projected areas $\pi a_gc_g/4$ and $\pi b_gc_g/4$: $A^\mathrm{eff}_p=\pi\sqrt{a_gb_g}c_g/4$, implying $A^\mathrm{eff}_p/A^\mathrm{max}_p=c_g/\sqrt{a_gb_g}\equiv S_f$, which is the Corey shape factor. Hence, we obtain $C_D=S_fC_{D\mathrm{Settle}}$ or
\begin{equation}
 C_D=4S_f(s-1)gd/(3v_s^2). \label{CdSettle}
\end{equation}

In the case of spherical grains ($S_f=1$), typical values for the quantities in (\ref{Thetat}) are $\mu_s\approx0.46$, $C_D\approx0.43$, and $\Theta_t\approx0.05$ \citep{Dealetal23a}, implying $c_t\approx0.03$. We use $c_t=0.032$ because it optimizes the agreement between the final model and the entire data compilation. On the other hand, when $s^{1/4}Ga\lesssim200$, viscous-sublayer effects render the scaling of $\Theta_t$ much more complicated \citep{Pahtzetal21}. Therefore, we use (\ref{Thetat}) to determine $\Theta_t$, except for the few numerical conditions with $s^{1/4}Ga<200$ (BLRANSx in Table~\ref{Data}, where `x' stands for an arbitrary character), for which we determine $\Theta_t$ directly from the simulations, as described by \citet{PahtzDuran20}. Note that the datasets J21DNS and KU17DNS in Table~\ref{Data}, which also satisfy $s^{1/4}Ga<200$, are excluded from this discussion as they violate the precondition $s^{1/2}Ga\gtrsim80$ required for (\ref{Q}).

\subsection{Extended data compilation} \label{DataCompilationExtended}
\subsubsection{Experiments and grain-resolved CFD-DEM simulations}
To evaluate our model, we require bedload flux datasets for shape-controlled grains, meaning that they must provide the values of $v_s$ and $S_f$, required to calculate $C_D$ in (\ref{CdSettle}), and the value of $\mu_s$, which affects $Q_\ast$ both via the calculation of $M_\ast$ in (\ref{M}) and via the calculation of $\Theta_t$ in (\ref{Thetat}). The datasets by \citet{Dealetal23a}, denoted as D23EXPx in Table~\ref{Data}, and by \citet{Zhangetal22,Zhangetal25}, denoted as Z22LESx and Z25LESx in Table~\ref{Data}, satisfy this requirement. However, we excluded from Z25LESx simulation data based on the ``artificial-shrinkage method'' by \citet{Zhangetal25}, since this method has been falsified \citep{Chenetal26}. Furthermore, in order to use the spherical-grain dataset NC18EXP by \citet{NiCapart18}, who did not report $\mu_s$, we acquired the same spheres from the same company, and measured $\mu_s$ using the funnel method \citep{BeakawiAlHashemiBaghabraAlAmoudi18} in the manner described by \citet{Dealetal23a}: by slowly pouring them onto an elevated disk bounded by a $2d_o$-high rim (Appendix~\ref{AOR}). Moreover, we scoured the literature and found two more usable experimental datasets for non-spherical grains: the bedload flux measurements for cylinders (R22EXPc) and lenses (R22EXPl) by \citet{Rebaietal22}. For both grain shapes, $v_s$ \citep{MatousekZrostlik20} and the grain dimensions (required for $S_f$) have been reported, while $\mu_s$ was measured using the funnel method in a box filled with water, which was sufficiently large to rule out potential box-sidewall friction effects (private correspondence with the authors). We confirmed with DEM simulations that different procedures of conducting the funnel method have only a small effect on $\mu_s$ ($<5\%$) provided that a sufficiently large number of grains is used and that there is some means of preventing grains from rolling away (such as a rim or rough bed; see Appendix~\ref{AOR}). By contrast, most other experimental datasets reported in the literature, even well-controlled ones \citep{Lajeunesseetal10}, do not report $S_f$ and/or $\mu_s$.

For those datasets mentioned in the previous paragraph that were not already discussed in \S\ref{DataCompilationSpherical} on the spherical-grain data compilation, we use the values of $h$ reported in the respective studies, as justified by $h^\ast=h/d\gg1$ (Table~\ref{Data}). For more details and the calculation of the bed shear stress uncertainty, see Appendix~\ref{BedShearStressEstimation}.

\subsubsection{Grain-unresolved CFD-DEM simulations}
Since our bedload flux model generalizes (\ref{M0}), the transport load expression by \citet{Bagnold56}, to (\ref{M}), involving the additional parameter $\overline{f_L}/\overline{f_D}=1.0$, we test it also for conditions to which (\ref{M0}) applies (i.e., $\overline{f_L}/\overline{f_D}=0$). For this purpose, we add data from grain-unresolved CFD-DEM simulations to the compilation, both from existing studies \citep{PahtzDuran20,Maurinetal18,Monthiller19} and from our own new simulations, all of which were conducted using either of the numerical models described by \citet{Duranetal12}, \citet{Maurinetal15} and \citet{Xieetal22} (Table~\ref{Data}; for details, see Appendix~\ref{UnresolvedSimulationDetails}). They differ from grain-resolved simulations in the fact that they do not resolve the flow around single grains, and they therefore incorporate empirical relations for the buoyancy, drag, and lift forces acting on grains, the latter of which are typically neglected (also here, i.e., $\overline{f_L}/\overline{f_D}=0$). They also allow us to perform thought experiments in which $C_D$ is treated as just another control parameter, since we can artificially modify $C_D$ through changing the drag force parameters (Appendix~\ref{UnresolvedSimulationDetails}); see LESCD and BLRANSCD in Table~\ref{Data}. Another purpose for adding these data is to confront our bedload flux model with transport conditions across a much larger range of the control parameters $\Theta$, $\alpha$, and $h^\ast$ than that spanned by the experiments and grain-resolved simulations alone.

\subsubsection*{Streamwise fluid momentum balance in BLRANSx}
The numerical model of \citet{Duranetal12} behind the datasets BLRANSx considers a flow in an inner turbulent boundary layer of infinite height (no sidewalls). To be consistent with (\ref{taubDefinitionAlt}), we slightly modified the streamwise fluid momentum balance employed by \citet{Duranetal12} to
\begin{equation}
 \d_z\sigma^f_{zx}=\rho_f\phi g\sin\alpha+\rho_p\phi f^{f\rightarrow p}_x,\quad\text{with}\quad\sigma^f_{zx}(\infty)=\tau_b, \label{MomentumDuran}
\end{equation}
before conducting simulations, where $\rho_p\phi\bm{f}^{f\rightarrow p}$ is the fluid--particle interaction force per unit volume of the mixture. In fact, integrating (\ref{MomentumDuran}) from zero to infinity, and combining the result with (\ref{IntMomSolidx}) and $\sigma_{zx}(0)=\sigma^f_{zx}(0)+\sigma^p_{zx}(0)$ leads precisely to (\ref{taubDefinitionAlt}). Note that, for the numerical model of \citet{Duranetal12}, the Shields number $\Theta\equiv\tau_b/(\rho_p\tilde g_zd)$ based on $\tau_b$ in (\ref{MomentumDuran}) is a prescribed fixed parameter.

\subsubsection*{Determination of $\Theta$ for M18RANS, M19RANSx, and LESx}
Since the raw data required to determine the bed surface elevation, and thus $h$ and $\tau_b$, are not available for the datasets M18RANS \citep{Maurinetal18} and M19RANSx \citep{Monthiller19} in Table~\ref{Data}, based on the model of \citet{Maurinetal15}, we infer $\Theta$ from their reported transport-driving Shields number $\Theta_\ast$, as explained in Appendix~\ref{ThetaUnresolved}.

\subsection{Model validation} \label{ModelValidation}
\begin{figure}
 \centering
 \includegraphics[width=1.0\columnwidth]{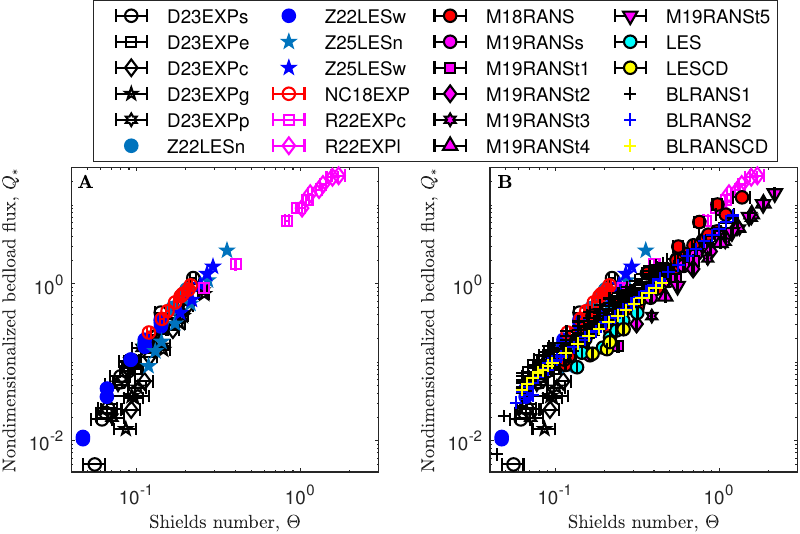}
 \caption{\textbf{Data compilation exhibits substantial scatter due to grain shape and other effects.} Non-dimensionalised bedload flux $Q_\ast$ versus Shields number $\Theta$, both rescaled using the generalized grain diameter definition $d\equiv\frac{3}{2}V_p/A^\mathrm{max}_p$ (see text). Symbols in A correspond to data from channel-geometry-corrected experiments and grain-resolved CFD-DEM simulations (Table~\ref{Data}). Symbols in B also include grain-unresolved CFD-DEM simulations (Table~\ref{Data}). An error bar indicates the standard error and/or uncertainty range. In the absence of error bars, uncertainties are smaller than the symbol size.}
 \label{DataRange}
\end{figure}

\begin{figure}
 \centering
 \includegraphics[width=1.0\columnwidth]{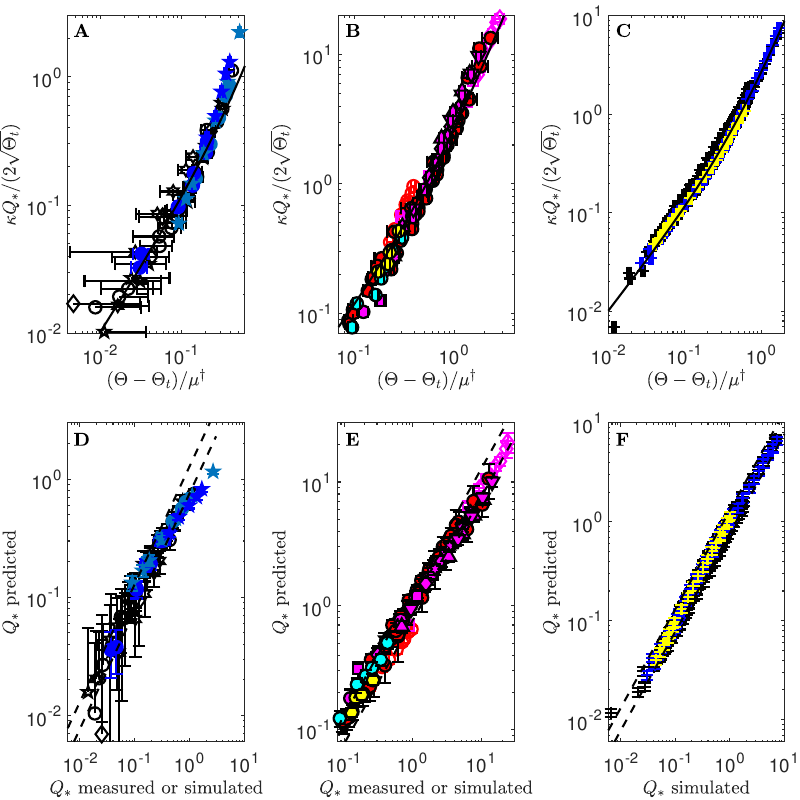}
 \caption{\textbf{Test of our rough-bedload flux model across grain shapes, bed slopes, flow strengths, and flow depths.} (A--C) Normalized bedload flux $Y\equiv\kappa Q_\ast/(2\sqrt{\Theta_t})$ versus $X\equiv(\Theta-\Theta_t)/\mu_\dagger$ and model prediction $Y=X(1+c_MX)$ (solid lines, from (\ref{Q}) and (\ref{M})). (D--F) Plots of $Q_\ast$ predicted by (\ref{Q}) and (\ref{M}) versus measured or simulated $Q_\ast$. Symbols correspond to channel-geometry-corrected data from a variety of methods (Table~\ref{Data}; see figure~\ref{DataRange} for legend). An error bar indicates the standard error and/or uncertainty range. In the absence of error bars, uncertainties are smaller than the symbol size. When $s^{1/4}Ga\geq200$, $\Theta_t$ is calculated by (\ref{Thetat}). Otherwise (only BLRANSx), $\Theta_t$ is determined directly from the simulations, as described in \citet{PahtzDuran20}. All grain-unresolved simulations neglect lift forces, implying $\overline{f_L}/\overline{f_D}=0$. Almost all data in D--F fall within a factor of $1.3$ of the model prediction, delineated by the dashed lines.}
 \label{ModelEvaluation}
\end{figure}

As a result of the wide range of control parameters and methods, especially grain-shape parameters, the data for $Q_\ast$ from the grain-unresolved simulations alone (figure~\ref{DataRange}A), and in extension from the entire data compilation (figure~\ref{DataRange}B), can vary by almost an order of magnitude at a given channel-geometry-corrected Shields number $\Theta$.
Nonetheless, the predictions from our bedload flux model deviate by less than a factor of $1.3$ from almost all these data (figure~\ref{ModelEvaluation}).
The only notable exceptions are the datasets Z25LESn and Z25LESw, which are based on the same grain shape. While the low-$\Theta$ simulations corresponding to these datasets are reproduced, their large-$\Theta$ simulations are substantially underpredicted (figures~\ref{ModelEvaluation}A,D). Among the other datasets, deviations of similar magnitude are seen only for two isolated data points: one for M19RANSt1 (figure~\ref{ModelEvaluation}E) and another one for BLRANS1 (figure~\ref{ModelEvaluation}F), both of which correspond to the smallest value of $\Theta$ for each respective dataset.

As a side note, the agreement between our bedload flux model and most of the data compilation also lends support to the reliability of the grain-unresolved simulations, since, if they were governed by different physics, then one would not expect them to be captured by the same physical analytical model.

\section{Discussion} \label{Discussion}
\subsection{The role of fluid lift in bedload transport}
Our rough-bedload flux model combines the global energy balance, (\ref{Q}) after \citet{PahtzDuran20}, linking the bedload flux to the transport load, with Bagnold's famous transport load expression \citep{Bagnold56}, (\ref{M}), generalized to account for a non-vanishing ratio $\overline{f_L}/\overline{f_D}$ between the average lift and drag forces acting on transported grains. This is curious in so far as fluid lift is often assumed to be negligible in bulk bedload transport \citep{PahtzDuran20,Pahtzetal21,Dealetal23a}, mainly because once a grain is elevated above a stationary bed, the shear-induced lift force acting on it rapidly declines with elevation \citep{Chepil61,Lietal19} and may even become negative \citep{Lietal19,ShiRzehak19}. In this regard, it is also interesting that $\overline{f_L}/\overline{f_D}$ has no bearing on the transport threshold $\Theta_t$ in (\ref{Thetat}). (We confirmed that including $\overline{f_L}/\overline{f_D}$ in (\ref{Thetat}) in a fashion after \citet{WibergSmith87} results in failure of the overall model.) The key difference between (\ref{M}) and (\ref{Thetat}) is that the former is modeling collective grain motion, whereas the latter is derived from the physical picture of a single test grain bouncing along an otherwise immobile bed. In DNS-DEM simulations, collective grain motion is highly associated with pressure-gradient-induced lift forces of a comparable magnitude to streamwise drag forces \citep{Jietal13}. By contrast, in DNS simulations of a fixed bed and a fixed particle elevated above it, such pressure-gradient-induced lift forces are much smaller than streamwise drag forces \citep{Lietal19}. (Note that \citet{Lietal19} called the pressure-gradient-induced lift forces ``wall-normal drag forces''.) This suggest that collective grain motion might induce pressure fluctuations and thereby substantially enhance lift forces, though more studies are needed to test this hypothesis and, if confirmed, elucidate the underlying physical mechanism.

\subsection{Test of the rough-bedload flux model of \citet{Dealetal23a} with independent data} \label{DealModelTest}
We are aware of only a single alternative bedload flux model attempting to account for grain shape: the one by \citet{Dealetal23a}. These authors generalized $d$ to the volume-equivalent sphere diameter $d_o$, as opposed to the shortest grain axis $c_g\approx S_f^{2/3}d_o$ in our model, and used the modified Shields number $\Theta^\mathrm{D23}\equiv\tau_b/(\rho_p\tilde g_zd_o)$ and modified non-dimensionalised bedload flux $Q^\mathrm{D23}_\ast\equiv Q/(\rho_pd_o\sqrt{(s-1)gd_o})$ to quantify the flow strength and transport rate, respectively. They proposed that bedload flux data corresponding to different grain shapes collapse onto a single curve $Q^\mathrm{D23}_\ast=f[(C^\ast/\mu^\ast)\Theta^\mathrm{D23}]$, where $C^\ast$ is the sphere-normalized value of $C_Dd_o/c_g$ (Appendix~\ref{DragDeal}) and $\mu^\ast\equiv(\mu_s-\tan\alpha)/(\mu_o-\tan\alpha)$, with $\mu_o=\tan24^\circ$ the approximate value of $\mu_s$ for spheres.

\citet{Dealetal23a} validated their model with their bedload flux measurements for several grain shapes, but not with other experimental data from independent sources.
\begin{figure}
 \centering
 \includegraphics[width=1.0\columnwidth]{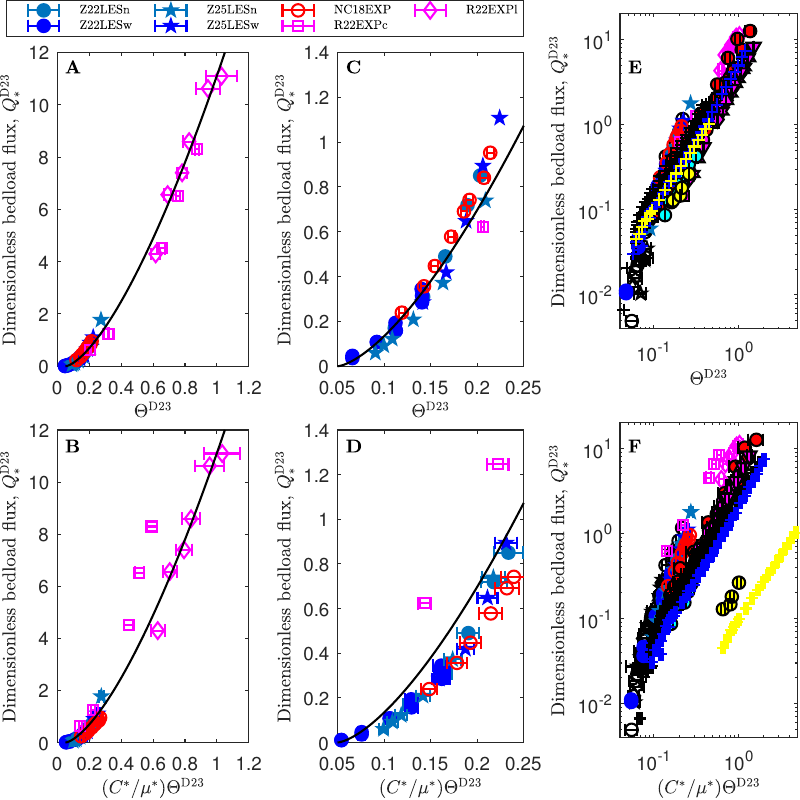}
 \caption{\textbf{Test of the bedload flux model of \citet{Dealetal23a} with independent data.} Plots of non-dimensionalised bedload flux $Q^\mathrm{D23}_\ast$ versus shape-uncorrected Shields number $\Theta^\mathrm{D23}$ (A,C,E) and shape-corrected Shields number $(C^\ast/\mu^\ast)\Theta^\mathrm{D23}$ (B,D,F), based on the same definitions as in \citet{Dealetal23a}. Symbols in (A--D) correspond to data from channel-geometry-corrected experiments and grain-resolved simulations from independent studies. Symbols in (E,F) correspond to the entire channel-geometry-corrected data compilation, including D23EXPx \citep{Dealetal23a} and data from grain-unresolved simulations (Table~\ref{Data}, see figure~\ref{DataRange} for legend). An error bar indicates the standard error and/or uncertainty range. In the absence of error bars, uncertainties are smaller than the symbol size. The solid lines correspond to $Q^\mathrm{D23}_\ast=12(\Theta^\mathrm{D23}-0.05)^{3/2}$ (A,C) and $Q^\mathrm{D23}_\ast=12(\Theta^\mathrm{D23}C^\ast/\mu^\ast-0.05)^{3/2}$ (B,D). The latter is the prediction by \citet{Dealetal23a}.}
 \label{DealFalsification}
\end{figure}
Here, we test their model with the independent datasets from our data compilation, sidewall-corrected with our derived universal method (figure~\ref{DealFalsification}). We remind the reader that, when applied to the datasets by \citet{Dealetal23a}, this correction yields almost the same bed shear stress $\tau_b$ as their own empirical correction (figure~\ref{SphereData2}A), while unifying spherical-grain datasets across experimental facilities (figure~\ref{SphereData}D). So, any unfavorable outcome of this test is likely not attributable to the application of our method. To be as fair as possible, we focus on those independent datasets obtained from experiments and grain-resolved simulations (figures~\ref{DealFalsification}A--D) and use the entire data compilation only as a secondary addition to our arguments (figures~\ref{DealFalsification}E,F). In addition, our main interest is not in the agreement between their model equation (solid lines in figures~\ref{DealFalsification}B,D) and the data, since this can be easily remedied through a change of the prefactor in that equation, but in the question of whether or not their Shields number correction reduces the scatter in the data.

It can be seen that the grain-shape correction by \citet{Dealetal23a} (figures~\ref{DealFalsification}B,D) much increases the scatter in the independent data when compared with the situation without correction (figures~\ref{DealFalsification}A,C), for both intense and weak transport conditions. In particular, the chosen range $0.05{-}0.25$ of the (un)corrected Shields number in figures~\ref{DealFalsification}C and \ref{DealFalsification}D is equivalent to the range of their own experiments \citep{Dealetal23a}, and the maximum scatter in figure~\ref{DealFalsification}D by approximately a factor of $3$ occurs well within this range, at $(C^\ast/\mu^\ast)\Theta^\mathrm{D23}\approx0.14$. It is mainly caused by the dataset R22EXPc in Table~\ref{Data}, for which $C^\ast/\mu^\ast=0.69$, a more extreme value of $C^\ast/\mu^\ast$ than in all their experiments ($C^\ast/\mu^\ast=0.84{-}1.05$) \citep{Dealetal23a}.

We believe that this poor performance chiefly results from their incorporation of $C^\ast$ in their grain-shape correction of the Shields number. In fact, it was shown that models that express bedload flux as a function of the excess bed shear stress are inherently linked to the assumption by \citet{Bagnold56}, later confirmed by \citet{PahtzDuran18b}, that the bed friction coefficient $\mu_b$ is constant. (In the model of \citet{Dealetal23a}, $\mu_b=\mu_s$, whereas $\mu_b=r_b\mu_s$ in our model.) In equilibrium, this constant friction coefficient is equal to the ratio between the average streamwise and bed-normal forces acting on transported grains, rendering the average drag force per unit mass $\overline{f_D}$ insensitive to the details of the flow. In consequence, increments of $C^\ast$ (or $C_D$ in our model) tend to be compensated by decrements of the average fluid--particle velocity difference, keeping $\overline{f_D}$ and therefore the effective transport-driving shear stress constant. Note that this reasoning is corroborated by the fact that the yellow symbols in figures~\ref{DealFalsification}E and \ref{DealFalsification}F, corresponding to conditions with artificially strongly elevated values of $C^\ast$ or $C_D$ (LESCD and BLRANSCD in Table~\ref{Data}), shift far away from the rest of the data after employing the grain-shape correction by \citet{Dealetal23a}. In contrast, they are captured by our bedload flux model (figure~\ref{ModelEvaluation}D) because it does not modify the Shields number $\Theta$ by a drag-dependent correction.

\section{Conclusions} \label{Conclusions}
We have derived a universal channel geometry correction method to determine the bed shear stress that collapses spherical-grain data of the non-dimensionalised bedload flux $Q_\ast$ as a function of the Shields number $\Theta$ (figure~\ref{SphereData}), obtained from experiments and grain-resolved CFD-DEM simulations, across a wide range of geometries, including narrow-deep and wide-shallow open channel flows. It consists of a granular-physics-based method to determine the bed surface elevation and a sidewall correction derived from Kolmogorov's theory of turbulence. The latter incorporates a single parameter ($c_{bw}\simeq2.23$ in (\ref{fwfb})) that is exclusively related to approximately channel-geometry-independent scaling coefficients in that theory. We have applied the channel geometry correction method to a very diverse compilation of grain-shape-controlled datasets from experiments, and grain-resolved and grain-unresolved CFD-DEM simulations (Table~\ref{Data}). An existing bedload flux model, here generalized to account for grain-shape effects, predicts almost all these data within a factor of $1.3$ (figure~\ref{ModelEvaluation}). By contrast, the only, to our knowledge, existing alternative bedload model accounting for grain-shape effects \citep{Dealetal23a} clearly disagrees with the data compilation (figure~\ref{DealFalsification}). While our model agrees with the latter model in that the Corey shape factor $S_f$, static angle of repose $\mu_s$, and settling drag coefficient $C_{D\mathrm{Settle}}$ are the three relevant parameters for modeling grain shape, the effect of $C_{D\mathrm{Settle}}$ on $Q_\ast$ is much weaker in our model, since it appears only via the transport threshold, and not also via a modification of $\Theta$. In \S\ref{DealModelTest}, we have reasoned that a modification of $\Theta$ via $C_{D\mathrm{Settle}}$ is unphysical, since it conflicts with the well-established result that the friction coefficient $\mu_b$ at the interface between bed and transport layer is constant \citep{Bagnold56,PahtzDuran18b}. Another important difference between the two models is the meaning of the grain diameter $d$ used to nondimensionalize physical quantities. \cite{Dealetal23a} assumed that $d$ is equal to the volume-equivalent grain diameter $d_o$, whereas we reasoned that $d$ should be the size $c_g\approx S_f^{2/3}d_o$ of the shortest grain axis.

In view of the diversity of the data compilation in terms of sediment transport and channel flow conditions, and the methods on which it is based on, the degree to which our relatively simple model, with only three adjustable parameters, agrees with it is remarkable (figure~\ref{ModelEvaluation}). This suggests that, with the here employed sidewall correction and bed surface definition, and our physical model to predict the flux of rough bedload transport (applicable if $s^{1/2}Ga\gtrsim80$, satisfied for most conditions in nature), bedload variability is largely diminished, at least for the idealized case considered here (steady flows over flat beds at short timescales). This is a prerequisite for discerning different sources of variability in the more complex situations typically encountered in nature.

\backsection[Acknowledgments]{The authors thank Yesheng Lu for helping to find grain-shape-controlled bedload transport datasets.}

\backsection[Funding]{Z.H. acknowledges financial support from the grant National Key R\&D Program of China (grant no. 2023YFC3008100). T.P. acknowledges financial support from the grants National Natural Science Foundation of China (grant nos.~12350710176, 12272344). O.D. acknowledges financial support from Texas A\&M Engineering Experiment Station.}

\backsection[Declaration of interests]{The authors report no conflict of interest.}

\backsection[Data availability statement]{The data required to reproduce figures~\ref{SphereData}, \ref{SphereData2}A, \ref{DataRange}, \ref{ModelEvaluation}, \ref{DealFalsification} and \ref{MeanVolumeFraction}, and a MATLAB code to plot them are available online at https://doi.org/10.5281/zenodo.20921059. The measurements of $\mu_s$ resulting from the experiment in figure~\ref{EXPAOR} and simulations in figure~\ref{DEMAOR} are reported in table~\ref{Data}. The data plotted in figure~\ref{SphereData2}B are from the tables of \citet{Guo15}. The data plotted in figure~\ref{NiCapartDeal} are from \citet{NiCapart18} and \citet{Dealetal23a} and can also be plotted using the MATLAB code. The data plotted in figures~\ref{BedSurfaceDetermination0} and \ref{BedSurfaceDeterminationSupport} are from the supplementary materials of \citet{NiCapart18}. All data needed to evaluate the conclusions of the paper are present in the paper.}

\backsection[Author contributions]{T.P. and Y.C. contributed equally.}

\appendix

\section{The Einstein--Johnson sidewall correction} \label{SidewallCorrectionEinstein}
The empirical sidewall correction by Einstein--Johnson calculates the friction factor ratio $f_w/f_b$ in (\ref{taubcorr}) as \citep{Guo17}
\begin{equation}
\begin{split}
 f_w/f_b&=R_w/R_b,\quad\text{with}\quad R_b=h(1-2R_w/b),\quad R_w=f^{-1}R[6\log_{10}\left[W(x)/x\right]]^{-2}, \\
 f&=8U^{-2}gR\sin\alpha,\quad x=[9\Rey/(100f)]^{1/3},
\end{split}\label{EinsteinJohnson}
\end{equation}
where $W(x)$ denotes the principal branch of the Lambert~$W$ function. These expressions constitute an analytical solution of the system of equations by \citet{Einstein42} and \citet{Johnson42}, improving their iterative solution method as well as the graphical approximation by \citet{VanoniBrooks57}.

\section{Estimation of bed shear stress and its uncertainty} \label{BedShearStressEstimation}

\subsection{Dataset NC18EXP}
For this highly-resolved dataset, the profiles of $\dot\gamma\Sigma\phi$ exhibit several local maximums due to layering (figure~\ref{BedSurfaceDetermination0}). We choose the top-most local maximum at the bed surface elevation, defined in (\ref{BedSurface}), because the profiles $\Theta^f_{zx}(z)$ exhibit a focal point there, with $\Theta^f_{zx}(0)\approx\Theta_t$ (figure~\ref{BedSurfaceDeterminationSupport}), consistent with the theoretical expectation \citep{PahtzDuran18b}. It occurs $\Delta z=31.75\pm0.5~\mathrm{mm}$ above the channel bottom for all flow conditions. The upper and lower bounds $\Delta z=32.25~\mathrm{mm}$ and $\Delta z=31.25~\mathrm{mm}$ then yield corresponding lower ($h_\mathrm{min}$) and upper ($h_\mathrm{max}$) bounds, respectively, of $h$ for a given flow condition, with corresponding bed shear stress estimates $\tilde\tau_b(h_\mathrm{max})$ and $\tilde\tau_b(h_\mathrm{min})$, respectively, obtained from (\ref{taubcorr}) and (\ref{fwfb}). Hence, we estimate $\tau_b$ and its uncertainty as
\begin{equation}
 \tau_b=\frac{1}{2}\left[\tilde\tau_b(h_\mathrm{max})+\tilde\tau_b(h_\mathrm{min})\right]\pm\frac{1}{2}\left[\tilde\tau_b(h_\mathrm{max})-\tilde\tau_b(h_\mathrm{min})\right],
\end{equation}

\subsection{Datasets D23EXPx, R22EXPx, Z22LESx, and Z25LESx}
For these datasets, we use the reported values of $h$ as its effect on $\tau_b$ is very small because $h$ was much larger than the bedload layer thickness \citep{Dealetal23a,Zhangetal22,Zhangetal25}. Note that, for the numerical datasets Z22LESx and Z25LESx, we therefore conclude that the uncertainty of $\tau_b$ is very small (smaller than the respective symbols in the plots). For the experimental datasets D23EXPx, the main source of uncertainty is the determination of the bed slope angle $\alpha$. We propagate its reported uncertainty to estimate the uncertainty of $\tau_b$. For the experimental datasets R22EXPx, the main source of uncertainty is the determination of $h$. We propagate its reported uncertainty to estimate the uncertainty of $\tau_b$.

\subsection{Datasets J21DNS and KU17DNS}
As we do not have access to the raw data of these datasets, and since $h$ was much larger than the bedload thickness \citep{Jainetal21,KidanemariamUhlmann17}, we use the reported values of $h^\ast$ and $\Theta$, assuming that the uncertainty of $\Theta$ is very small (smaller than the respective symbols in the plots).

\subsection{Datasets M18RANS, M19RANSx, LESx}
For these datasets, see Appendix~\ref{ThetaUnresolved}.

\subsection{Datasets BLRANSx}
For these datasets, $h^\ast=\infty$ by construction, and $\Theta$ is a prescribed parameter (no uncertainty; see \S\ref{DataCompilationExtended}).

\section{Transport load expression} \label{TransportLoadExpression}
To derive simple expressions for the dimensionless transport load $M_\ast$, we neglect the contributions of particle-sidewall interactions in the streamwise and vertical particle momentum balances. Then, these balances read \citep{Pahtzetal26}
\begin{align}
 \d_z\sigma^p_{zx}=-\rho_p\phi g\sin\alpha-\rho_p\phi f^{f\rightarrow p}_x, \label{MomSolidx} \\
 \d_z\sigma^p_{zz}=\rho_p\phi g\cos\alpha-\rho_p\phi f^{f\rightarrow p}_z, \label{MomSolidz}
\end{align}
where $\rho_p\phi\bm{f}^{f\rightarrow p}$ is the fluid--particle interaction force per unit volume of the mixture. In both balance equations, we decompose $\rho_p\phi\bm{f}^{f\rightarrow p}$ into buoyancy ($\phi\d_z\partial_i\sigma_{ix}$) and non-buoyancy contributions, where the latter are identified as the drag plus pressure gradient ($\rho_p\phi f_D+\phi\chi$) and lift ($\rho_p\phi\tilde f_L$) forces per unit volume of the mixture for the streamwise and vertical directions, respectively:
\begin{align}
 f^{f\rightarrow p}_x&=f_D+\chi/\rho_p+\d_z(\sigma_{zx}-\sigma_{\mathrm{Re}zx})/\rho_p\simeq f_D+\chi/\rho_p, \label{fx} \\
 f^{f\rightarrow p}_z&=\tilde f_L+\d_z(\sigma_{zz}-\sigma_{\mathrm{Re}zz})/\rho_p\simeq\tilde f_L+\d_z\sigma_{zz}/\rho_p=\frac{\tilde f_L}{1-\phi}+s^{-1}g\cos\alpha, \label{fz}
\end{align}
The buoyancy contribution in (\ref{fx}) has been neglected, since the effective fluid shear stress $\sigma_{zx}$ is dominated by the Reynolds stress $\sigma_{\mathrm{Re}zx}$, gradients of which do not contribute to buoyancy \citep{Maurinetal18,Zhuetal26}. By contrast, the effective fluid pressure $\sigma_{zz}$ dominates the Reynolds pressure $\sigma_{\mathrm{Re}zz}$, leading to a standard buoyancy expression \citep[e.g.]{Duranetal12}. Using (\ref{fx}) and (\ref{fz}), integration of (\ref{MomSolidx}) and (\ref{MomSolidz}), respectively, yields
\begin{align}
 \sigma^p_{zx}(0)&=M\left(g\sin\alpha+\chi/\rho_p+\overline{f_D}\right), \label{IntMomSolidx} \\
 \sigma^p_{zz}(0)&=-M\left(\tilde g_z-\overline{f_L}\right), \label{IntMomSolidz}
\end{align}
where $f_L\equiv\tilde f_L/(1-\phi)$. It follows that the bed friction coefficient, $\mu_b\equiv-\sigma^p_{zx}(0)/\sigma^p_{zz}(0)$, constrains the average forces acting on transported grains:
\begin{equation}
 \mu_b=\frac{g\sin\alpha+\chi/\rho_p+\overline{f_D}}{\tilde g_z-\overline{f_L}}. \label{mub}
\end{equation}
This equation can be transformed into
\begin{equation}
 \overline{f_D}=\frac{\mu_b\tilde g_z-g\sin\alpha-\chi/\rho_p}{1+\mu_b\overline{f_L}/\overline{f_D}}. \label{Fxsurf}
\end{equation}
Finally, combining (\ref{IntMomSolidx}) and (\ref{Fxsurf}) with (\ref{taubDefinitionAlt}) and $\sigma_{zx}=\sigma^f_{zx}+\sigma^p_{zx}$, nondimensionalizing, and employing $\Theta-\Theta^f_{zx}(0)\equiv(\tau_b-\sigma^f_{zx}(0))/(\rho_p\tilde g_zd)=\Theta-\Theta_t$ \citep{PahtzDuran18b} leads to (\ref{M}).

\section{Determinations of angle of repose and its uncertainty} \label{AOR}
\begin{figure}
 \centering
 \includegraphics[width=0.5\columnwidth]{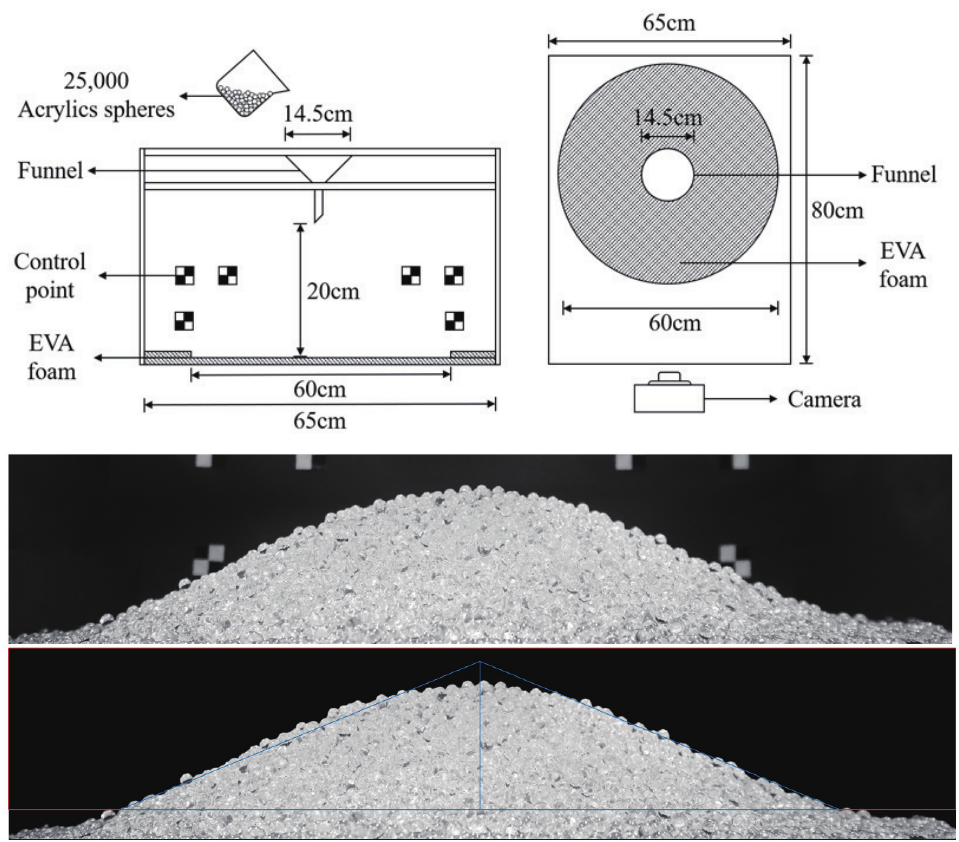}
 \caption{\textbf{Measurements of $\mu_s$ for the dataset NC18EXP (Table~\ref{Data}).} We slowly poured $25,000$ grains onto an elevated disk bounded by a $2d$-high rim. The angle of repose of the resulting heap is then determined as the base angle of the isosceles triangle that has the same area as the projected heap, averaged over two side-view images separated by a rotation of $90^\circ$ \citep{ElekesParteli21}.}
 \label{EXPAOR}
\end{figure}

\begin{figure}
 \centering
 \includegraphics[width=1.0\columnwidth]{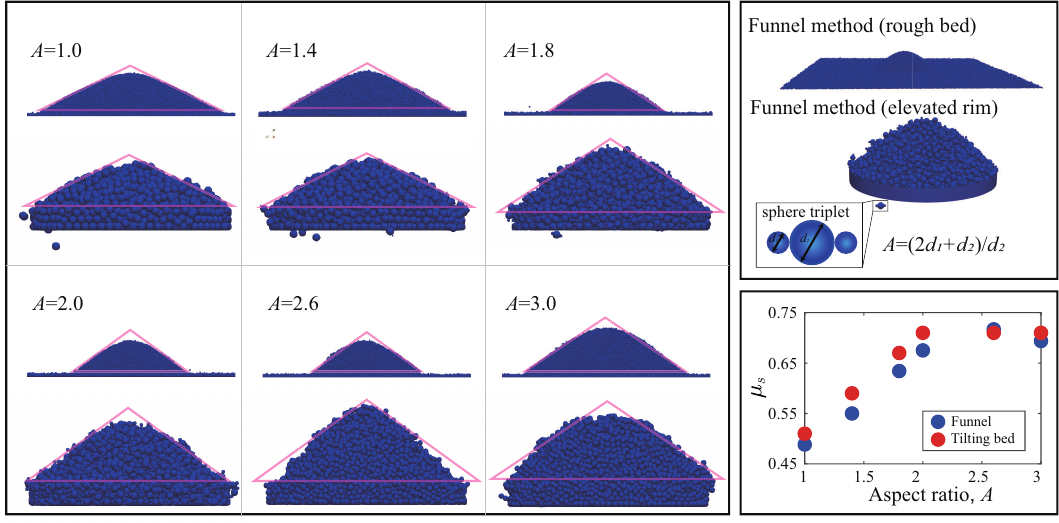}
 \caption{\textbf{Numerical determination of the static angle of repose.} We compare three different methods of measuring the tangent of the static angle of repose, $\mu_s$, for the composite grains (spheres and sphere triplets) used in the grain-unresolved DEM-CFD simulations: slowly pouring a sufficient number of grains onto a rough bed consisting of spheres of the same volume-equivalent sphere diameter $d_o$ (funnel method~1); onto an elevated disk of diameter $24d_o$ bounded by a $2d_o$-high rim (funnel method~2, exactly as described in \citet{Dealetal23a}), and tilting the granular bed until the granular bulk moves (tilting method). The most reliable and reproducible results are yielded by funnel method~1, where the angle of repose is determined as the base angle of the isosceles triangle that has the same area as the projected heap, averaged over two side-view images separated by a rotation of $90^\circ$ \citep{ElekesParteli21}. For funnel method~2, the heaps exhibit too irregular shapes when the aspect ratio $A$ becomes too large, making a clear determination of the angle of repose difficult. However, for $A\lesssim2$, it yields values very close to funnel method~1. The tilting method also yields similar results to funnel method~1, though it is sometimes difficult to distinguish between bulk and isolated-grain motion. We therefore decided to use funnel method~1 as the standard procedure to determine $\mu_s$ in the grain-unresolved DEM-CFD simulations.}
 \label{DEMAOR}
\end{figure}

We acquired the same spheres as those used in \citet{NiCapart18} (NC18EXP in Table~\ref{Data}) from the same company (Chiao Dar Acry \& Advertisement Co., Ltd., \url{http://www.bridgeacry.com.tw}) and measured $\mu_s$ using the funnel method \citep{BeakawiAlHashemiBaghabraAlAmoudi18} in the manner described in \citet{Dealetal23a} (figure~\ref{EXPAOR}).
Furthermore, we confirmed with DEM simulations that different measurement methods have only a small effect on $\mu_s$ provided that a sufficiently large number of grains is used and that there is some means of preventing grains from rolling away (such as a rim or a rough bed, see figure~\ref{DEMAOR}).
Based on these simulations, we assign an uncertainty of $5\%$ to all $\mu_s$-measurements, including those from different studies and methods.

\section{Grain-unresolved simulation details} \label{UnresolvedSimulationDetails}
\subsection{Contact parameters and $\mu_s$ in grain-unresolved CFD-DEM simulations} \label{ContactParameters}
The normal restitution coefficient $e_n$, tangential contact friction coefficient $\mu_c$, and tangent of the static angle of repose, $\mu_s$ (determined as described in figure~\ref{DEMAOR}), are $(e_n,\mu_c,\mu_s)=(0.5,0.4,0.40)$ for the model of \citet{Maurinetal15} and data of \citet{Maurinetal18}, $(e_n,\mu_c,\mu_s)=(0.5,0.5,0.49{-}0.72)$ for the model of \citet{Maurinetal15} and data of \citet{Monthiller19}, $(e_n,\mu_c,\mu_s)=(0.3,0.5,0.49)$ for the model of \citet{Xieetal22}, and $(e_n,\mu_c,\mu_s)=(0.9,0.5,0.38)$ for the two-dimensional model of \citet{Duranetal12}, the latter of which exhibits a lower value of $\mu_s$ because the simulation domain was quasi-two-dimensional (all grains exhibited the same fixed lateral position, $y=0$).

\subsection{Drag coefficient}
The numerical CFD-DEM models that do not resolve the sub-grain scale rely on semiempirical relations for fluid--particle interactions. They consider buoyancy and drag forces on spheres or composite grains consisting of non-overlapping component spheres, but neglect lift forces. Following \citet{Sunetal17}, the drag force is modeled as the total force on all component spheres:
\begin{equation}
 \bm{F_D}=\sum_i\frac{1}{8}\rho_f\pi d_i^2\left[\left(\frac{\Rey_c\nu}{d_i\epsilon^{p_1}}\right)^{1/m}+\left(\frac{C^\infty_D|\bm{u_r}|}{\epsilon^{p_2}}\right)^{1/m}\right]^m\bm{u_r}, \label{FD}
\end{equation}
where $d_i$ is the diameter of the $i$-th component sphere, $\bm{u_r}$ is the fluid--particle velocity difference, and $\Rey_c$, $C^\infty_D$, $m$, $p_1$, and $p_2$ are empirical parameters. They are $(\Rey_c,C^\infty_C,m,p_1,p_2)=(24.4,0.4,1,3.1,3.1)$ in \citet{Maurinetal15}, $(\Rey_c,C^\infty_C,m,p_1,p_2)=(24,0.5,2,0,0)$ in \citet{Duranetal12} (nominal), $(\Rey_c,C^\infty_C,m,p_1,p_2)=(96,2,2,0,0)$ in \citet{Duranetal12} (enhanced drag), $(\Rey_c,C^\infty_C,m,p_1,p_2)=(23.04,0.3969,2,p,p-1)$ in \citet{Xieetal22} (nominal), and $(\Rey_c,C^\infty_C,m,p_1,p_2)=(69.12,1.1907,2,p,p-1)$ in \citet{Xieetal22} (enhanced drag), with $p=3.7-0.65\exp[-(1.5-\log_{10}(\epsilon|\bm{u_r}|d_i/\nu))^2/2]$. Dividing (\ref{FD}) by the grain weight $\frac{\pi}{6}\sum_id_i^3$ leads to an expression for the ratio between the drag acceleration $\bm{a_D}$ and submerged gravitational acceleration $\tilde g\equiv(1-1/s)g$ in terms of non-dimensionalised quantities:
\begin{equation}
 \frac{\bm{a_D}}{\tilde g}=\frac{3}{4}\left[\left(\frac{\Rey_c}{\widetilde{Ga}\epsilon^{p_1}}\right)^{1/m}+\left(\frac{C^\infty_D|\bm{{\tilde u_r}}|}{\epsilon^{p_2}}\right)^{1/m}\right]^m\bm{{\tilde u_r}}, \label{aD}
\end{equation}
where $\bm{{\tilde u_r}}\equiv\bm{u_r}/\sqrt{s\tilde gd}$, with $d\equiv\frac{3}{2}V_p/A^\mathrm{max}_p=\sum_id_i^3/\sum_id_i^2$, is the non-dimensionalised fluid--particle velocity difference and
\begin{equation}
 \widetilde{Ga}\equiv\frac{\sqrt{\left(\sum_id_i^2\right)\left(\sum_id_i^3\right)}}{\sum_id_i}\sqrt{s\tilde g}/\nu \label{Ga}
\end{equation}
a quantity similar to the Galileo number. From (\ref{aD}), one obtains the non-dimensionalised settling velocity $\tilde v_s$ of a single composite grain in quiescent fluid ($\epsilon=1$) as \citep{Pahtzetal21}
\begin{equation}
 \tilde v_s=\left[\sqrt{\frac{1}{4}\sqrt[m]{\left(\frac{\Rey_c}{C^\infty_D\widetilde{Ga}}\right)^2}+\sqrt[m]{\frac{4}{3C^\infty_D}}}-\frac{1}{2}\sqrt[m]{\frac{\Rey_c}{C^\infty_D\widetilde{Ga}}}\right]^m \label{vs}
\end{equation}
and subsequently the drag coefficient as
\begin{equation}
 C_D=\frac{4(s-1)gd}{3v_s^2}=\frac{4}{3\tilde v_s^2}.
\end{equation}
Here we used that a composite grain is treated as if the projected area of all its component spheres is always, regardless of its orientation relative to the flow, fully exposed to the flow, which follows from (\ref{FD}). That is, the total projected area seen by the flow, $A_p=\frac{\pi}{4}\sum_id_i^2$, is constant, and therefore $A^\mathrm{eff}_p=A^\mathrm{max}_p$ and $C_D=C_{D\mathrm{Settle}}$.

\section{Determination of $\Theta$ for M18RANS, M19RANSx, and LESx} \label{ThetaUnresolved}
For the datasets M18RANS \citep{Maurinetal18} and M19RANSx \citep{Monthiller19} in Table~\ref{Data}, based on the model of \citet{Maurinetal15}, only the transport-driving Shields number $\Theta_\ast$ is reported. \citet{Maurinetal15} defined $\Theta_\ast\equiv\tau_\mathrm{cl}/(\rho_p\tilde g_zd)$ as the Shields number based on the clear-water fluid shear stress $\tau_\mathrm{cl}$ on the top of the bedload layer:
\begin{equation}
 \tau_\mathrm{cl}\equiv\max\sigma^f_{zx}(z). \label{taucl}
\end{equation}
Using $\tau_b\propto h$ for their sidewall-free channels, $\tau_\mathrm{cl}$ therefore defines an effective clear-flow depth as
\begin{equation}
 h_\mathrm{cl}\equiv h\tau_\mathrm{cl}/\tau_b
\end{equation}
and subsequently an effective bedload layer thickness above the bed surface as
\begin{equation}
 h_p\equiv h-h_\mathrm{cl}. \label{hp}
\end{equation}
The latter is linked to the transport load $M$ via
\begin{equation}
 M=\rho_ph_p\underline{\phi},
\end{equation}
with $\underline{\phi}\equiv\frac{1}{h_p}\int_0^h\phi\d z\simeq\frac{1}{h_p}\int_0^{h_p}\phi\d z$ the approximate average of the particle volume fraction $\phi$ over the bedload layer thickness. Combining the above relations and nondimensionalizing the result yields
\begin{equation}
 \Theta=\Theta_\ast\left(1+\frac{M_\ast}{\underline{\phi}h^\ast_\mathrm{cl}}\right), \label{Thetaast}
\end{equation}
where $h^\ast_\mathrm{cl}\equiv h_\mathrm{cl}/d$ is the dimensionless clear-flow depth.

\begin{figure}
 \centering
 \includegraphics[width=0.5\columnwidth]{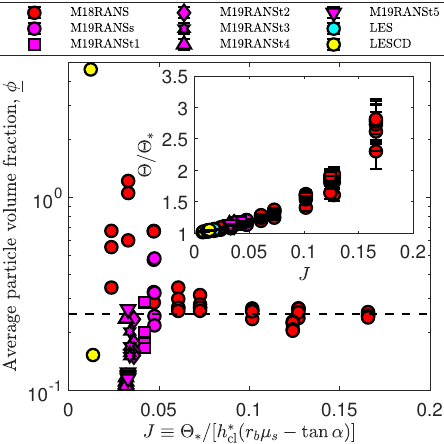}
 \caption{\textbf{Determination of transport-layer-averaged particle volume fraction $\underline{\phi}$.} $\underline{\phi}$, estimated from the $Q_\ast$-data as described in the text, versus $J$. Symbols correspond to grain-unresolved CFD-DEM simulations (Table~\ref{Data}). The dashed line indicates the value $\underline{\phi}=0.25$ exhibited for large $J$, where the ratio $\Theta/\Theta_\ast$ deviates from unity the most. Inset: Estimate of $\Theta/\Theta_\ast$ resulting from $\underline{\phi}=0.25$, where the error bars correspond to the propagated uncertainty of $\mu_s$, required for predicting $M_\ast$ and therefore $\Theta$, as described in the text.}
 \label{MeanVolumeFraction}
\end{figure}

To determine the parameter $\underline{\phi}$ in (\ref{Thetaast}), we combine (\ref{Thetaast}) with (\ref{M0}), valid for the grain-unresolved simulations due to $\overline{f_L}/\overline{f_D}=0$, yielding
\begin{equation}
 \underline{\phi}=J\left(1-\frac{\Theta_\ast-\Theta_t}{(r_b\mu_s-\tan\alpha)M_\ast}\right)^{-1}, \label{phiEstimate}
\end{equation}
with $J\equiv\Theta_\ast/[h^\ast_\mathrm{cl}(r_b\mu_s-\tan\alpha)]$. The unknown values of $M_\ast$ in (\ref{phiEstimate}) can be estimated from the simulation data of $Q_\ast$ via (\ref{Q}), using the predicted values of $\Theta_t$ from (\ref{Thetat}). Figure~\ref{MeanVolumeFraction} shows the resulting \textit{estimates} of $\underline{\phi}$ as a function of $J$ for the grain-unresolved datasets M18RANS, M19RANSx, LES, and LESCD.
Note that the estimates for $J\lesssim0.05$ cannot be trusted, since they are very sensitive to even slight variations in the simulation data of $\Theta_\ast$, $h^\ast_\mathrm{cl}$, and especially $Q_\ast$, explaining the large scatter and the occasionally unphysically large values of $\underline{\phi}$ for low $J$. However, this is not a serious issue, since for weak transport conditions, $\Theta_\ast$ is close to $\Theta$ no matter the precise value of $\underline{\phi}$ as $M_\ast$ in (\ref{Thetaast}) becomes small. By contrast, for $J\gtrsim0.05$, the estimations of $\underline{\phi}$ from the simulation data scatter relatively little between only approximately $0.2$ and $0.3$ (figure~\ref{MeanVolumeFraction}). In particular, for the most intense conditions, $\underline{\phi}\simeq0.25$, which is a reasonable value as it is approximately half of $\phi(0)$ for intense transport conditions \citep{PahtzDuran18b}. We therefore fix $\underline{\phi}=0.25$ in (\ref{Thetaast}) as a simple means to relate $\Theta$ to $\Theta_\ast$ for arbitrary conditions through (\ref{Thetaast}). The ratio $\Theta/\Theta_\ast$ resulting from this choice is shown in the inset of figure~\ref{MeanVolumeFraction}.

Although we have used the $M_\ast$-values estimated from the $Q_\ast$-data via (\ref{Q}) to obtain $\underline{\phi}=0.25$, we do not do so when estimating $\Theta$ from (\ref{Thetaast}), in order to avoid an implicit dependence of $\Theta$ on $Q_\ast$. Instead, we obtain $\Theta$ from the values of $M_\ast$ predicted by (\ref{M}). The uncertainty of $\Theta$ then corresponds to the propagated uncertainty of $\mu_s$ in the calculation of $\Theta$ via (\ref{M}) and (\ref{Thetaast}).

\section{Sphere-normalized drag coefficient after \citet{Dealetal23a}} \label{DragDeal}
\citet{Dealetal23a} defined the sphere-normalized drag coefficient $C^\ast$ as
\begin{equation}
 C^\ast\equiv S_fv_o^2/v_s^2,
\end{equation}
where $v_o$ is the theoretical value of the settling velocity of spheres, given via \citep{Dietrich82a,Dealetal23a}
\begin{equation}
 v_{o\ast}=-3.81564+1.94593D_\ast-0.09016D_\ast^2-0.00855D_\ast^3+0.00075D_\ast^4,
\end{equation}
with $v_{o\ast}\equiv\log_{10}v_o^3/[(s-1)g\nu]$ and $D_\ast\equiv\log_{10}(s-1)gd_o^3/\nu^2$. Since these expressions are based on the volume-equivalent sphere diameter $d_o$, as opposed to the shortest grain axis $c_g$ in our model, $C^\ast$ is linked to the drag coefficient $C_D$ in (\ref{CdSettle}) through $C^\ast=C_Dd_o/(C_oc_g)$, where $C_o\equiv4(s-1)gd_o/(3v_o^2)$.

\bibliographystyle{jfm}

\end{document}